\documentstyle[11pt,doublespace,margins,taus,epsf]{report}

\begin{document}

\pagenumbering{roman}
\pagestyle{plain}

\title{Postsynaptic mechanisms for the induction of long-term potentiation
and long-term depression of synaptic transmission in
CA1 pyramidal neurons of the hippocampus\\ \vspace*{.5in}
A Dissertation Submitted to the Faculty of \\
The Graduate School \\
Baylor College of Medicine \\
In Partial Fulfillment of the\\ 
Requirements for the Degree\\ of \\
Doctor of Philosophy }
\author{Lalania Kaye Schexnayder\\
Houston, Texas \\
August 25, 1999 }
\date{}

\maketitle

\chapter*{Abstract}
\addcontentsline{toc}{chapter}{Abstract}
\setcounter{page}{1} 

		The studies described in this dissertation have attempted to address the cellular mechanisms of information storage by the brain.  My work has focused primarily on the postsynaptic events that occur during and after induction of long-term potentiation (LTP) and long-term depression (LTD) at the Schaffer collateral-CA1 synapse of the hippocampus.  I have explored various aspects of the role of postsynaptic Ca$^{2+}$ in the induction of LTP and LTD using both extracellular and whole-cell recording techniques, as well as fluorescence imaging of Ca$^{2+}$.  I have also examined the possible role of modulation of dendritic K$^{+}$ channels in the induction of LTP.

	Increases in postsynaptic Ca$^{2+}$ are known to play an important part in the induction of both LTP and LTD, but the relative roles of different sources of Ca$^{2+}$ are not yet well delineated.  I have attempted to address the contributions of different sources of postsynaptic Ca$^{2+}$ in the induction of homosynaptic LTD.  My experiments have shown that Ca$^{2+}$ influx via at least two subtypes of voltage-gated Ca$^{2+}$ channel (VGCC) appears to play an important role in the induction of homosynaptic LTD in the CA1 region in vitro.  These include the high-threshold L-type channel as well as the low-threshold T-type channel.  

	Although it is well established that increases in postsynaptic Ca$^{2+}$ give rise to both LTP and LTD, it is still of interest how this shared mechanism can give rise to two opposing processes.  I have attempted to address the mechanisms by which the magnitude and direction of changes in synaptic strength are determined by the size of the increase in postsynaptic Ca$^{2+}$ and, specifically, the size of the influx via the NMDA receptor and the L-type voltage-gated Ca$^{2+}$ channel during plasticity-inducing stimulation.  In experiments utilizing whole-cell recordings in combination with fluorescence imaging of postsynaptic Ca$^{2+}$, I demonstrated that the size of the increase in postsynaptic Ca$^{2+}$ achieved during plasticity-inducing stimulation varies with frequency of stimulation and appears to determine the size and direction of changes in synaptic strength.  LTD was achieved with stimulation from 3 to 10 Hz while LTP was achieved with stimulation of 30 Hz and above.  Although there was no precise magnitude of postsynaptic Ca$^{2+}$ signal that corresponded to the division between LTP and LTD, there was a range of Ca$^{2+}$ values above which LTP was observed and below which LTD was observed.  In these experiments both nimodipine and APV blocked Ca$^{2+}$ signals to a similar extent and shifted the crossover point of the plasticity vs. frequency curve toward higher frequencies of stimulation, such that frequencies that normally resulted in LTP resulted in LTD, and frequencies that normally resulted in LTD gave rise to no plasticity. Thus, both voltage-gated Ca$^{2+}$ channels and NMDA receptors appear to contribute to the Ca$^{2+}$ influx necessary for the induction of synaptic plasticity. In addition, APV appears to block plasticity without causing a reduction in somatic Ca$^{2+}$ influx, suggesting that increases in dendritic [Ca$^{2+}$]$_{i}$ are more relevant for the induction of plasticity than those in the soma.

          The apical dendrites of CA1 hippocampal pyramidal neurons have been shown to contain a large density of transient or A-type K$^{+}$ channels which are known to play an important role in the modulation of the size of EPSPs and back-propagating action potentials.  In addition, these channels have been shown to be a substrate for  phosphorylation by several kinases implicated in LTP.  Upon addition of activators of these protein kinases, the activity of the channels is downregulated, resulting in an increase in neuronal excitability.  In light of these data, we hypothesized that LTP induction might result in localized decreases in dendritic K$^{+}$ channel activity and increases in amplitude of back-propagating action potentials, and proposed to observe these increases in amplitude indirectly as localized increases in dendritic Ca$^{2+}$ signals following LTP induction.  Application of 100 Hz pairing stimulation resulted in LTP induction and indeed gave rise to consistent localized increases in the size of dendritic Ca$^{2+}$ signals 5 min after stimulation.  These increases were demonstrated to occur within the 25$\mu$m segment of the dendrite shown to correspond to the site of synaptic input, or in the 25$\mu$m segment adjacent to it.  Overall strong depolarization of the dendritic tree did not appear to be sufficient to activate the mechanisms responsible for the localized increases in dendritic Ca$^{2+}$ signals.  Rather, it appears that there is also a need for synaptic input.  Increases in the size of the dendritic Ca$^{2+}$ signals were blocked by 50$\mu$M APV and the broad spectrum kinase inhibitor H7, both of which also blocked LTP.  The PKA inhibitor H89 did not block the induction of LTP but blocked the localized increases in dendritic Ca$^{2+}$ signals normally observed following LTP, suggesting that PKA is necessary to give rise to the localized increases in dendritic Ca$^{2+}$, although it is not necessary for the induction of the early phase of LTP.  We concluded that the high-frequency stimulation (HFS) used to induce LTP results in localized increases in dendritic Ca$^{2+}$ influx which may represent a localized increase in action potential amplitude at or near the synapse.  We propose that this occurs through the activation of one or more kinases which act locally at the synapse to phosphorylate dendritic K$^{+}$ channels and to decrease their activity.  

\chapter*{Acknowledgments}

\addcontentsline{toc}{chapter}{Acknowledgments}

I would like to express my sincere gratitude to the many individuals who have provided help and encouragement throughout my years of graduate school in the Division of Neuroscience.  First I would like to thank my advisor, Dan Johnston.    His expertise in electrophysiology and his stringent experimental standards helped to create an environment in which I felt continually challenged and where I learned to utilize the most recent techniques in cellular neurophysiology.  He has always fostered an open environment for scientific discussion and has always been available to discuss my experimental successes and failures, providing encouragement and support through the many changes my projects have undergone.  It has been a privelege to work with him and to be a member of his lab.

I would like to thank my committee members, Ruth Anne Eatock, Peter Saggau, John Swann, and David Sweatt.  Throughout my years of graduate school they have been a source of support and valuable advice, helping to guide both my experimental plans and my development as a scientist.  Their doors were always open to discuss my progress and to help me plan future experiments.  I have truly enjoyed working with all of them.  

I am particularly indebted to Rick Gray, as are all members of the Johnston lab, for being a constant and reliable source of information and assistance regarding all things technical, electronic, and computer related.  He has never failed to help me with whatever problems presented themselves, whether trivial or catastrophic, and I believe I have learned more from Rick than from any other member of the lab.  He has also been a great friend, and our daily discussions regarding everything from neuroscience to astrology have made my time in the lab much more enjoyable.

I would also like to thank my fellow lab members.  Costa Colbert provided data acquisition and analysis software that I have used throughout my years in the lab, and he was a tremendous help to me in learning electrophysiology techniques in my early days in the lab.  I would like to thank Brian Christie with whom I collaborated on calcium imaging experiments, and Dax Hoffman on whose data some of my experiments are based.  Other Johnston lab members with whom I have enjoyed working include Ajay Kapur, Bob Avery, Erik Cook, Jeff Magee, Mark Yeckel, Nick Poolos, Shigeo Watanabe, and Li-lian Yuan.  I am also endebted to Mahmud Haque for his calcium imaging analysis software and frequent computer assistance.  

I would like to thank the Mary Kennedy lab at Caltech with whom I collaborated on the CaMKII experiments.  I would like to thank my fellow graduate students in the Division of Neuroscience, and all members of the Division.  It has been a pleasure and a privelege to have done my graduate studies in this environment.  I would like to thank the many members of the administrative staff in the Division who have made my life easier, including Debbie Colbert, Vicki Cox, Donna Campbell and Diane Jensen.  I would like to thank the MD/PhD Program, in particular Dr. Jim Lupski and Dr. Marty Matzuk, as well as Kathy Crawford and Suzanne Barnes.  

Lastly, I would like to thank my husband and best friend of many years, Shane Guillory.  He has been a constant source of love and support throughout my lengthy education, and has sacrificed much so that I could pursue my dreams and ambitions.  I am endebted to him most of all.      

\tableofcontents
\listoffigures
\addcontentsline{toc}{chapter}{List of Figures}

\pagestyle{plain}

\chapter*{1 Introduction and Background}
\setcounter{chapter}{1}
\addcontentsline{toc}{chapter}{1 Introduction and Background} 
\label{intro}

\pagenumbering{arabic}

	As human beings with a conscious existence and self-awareness, we are perhaps cursed with the ability to contemplate the mind. Struggles to understand the seat of consciousness have often led to the study of memory, perhaps the most concrete manifestation of the work that the mind performs. Disorders of memory and cognition have provided neuroscientists many clues to the biological mechanisms by which the brain stores and processes information. In 1957, a patient known only by the initials HM underwent bilateral removal of the region of the brain known as the hippocampus for the treatment of intractable epilepsy, leaving him unable to encode new information into long-term memory. The neuropsychological studies of this patient established a critical link between the hippocampus and the formation of new memories [.scoville milner @16646@.] sparking the interest of neuroscientists and making the hippocampus one of the most studied regions of the brain in the quest to understand learning and memory.

	In the experiments described in this dissertation, I have investigated the phenomenon known as synaptic plasticity, the means by which connections between neurons are thought to be strengthened or weakened in response to neuronal activity.  Specifically, I have studied the phenomena of hippocampal long-term potentiation (LTP) and long-term depression (LTD) in the pyramidal neurons of the CA1 region.  My interest has been primarily in the physiological and biochemical events that occur in the postsynaptic neuron during and after the induction of LTP and LTD.  My experiments have focused on the involvement of various sources of postsynaptic Ca$^{2+}$ in regulating the magnitude and direction of changes in synaptic strength, as well as on the mechanism by which modulation of dendritic transient or A-type K$^{+}$ channels may participate in the induction of LTP. 

\section{Long-term potentiation and long-term depression}

Use-dependent modification of synaptic efficacy has been observed in various nervous tissues and can result in lasting potentiation or depression of synaptic transmission.  Long-term potentiation (LTP) and long-term depression (LTD) are forms of synaptic plasticity thought to be involved in learning and memory and which provide an attractive cellular mechanism for the storage of information by the brain.  

	Long-term potentiation was first studied in the hippocampus at the synapse of the perforant path and the granule cells of the dentate gyrus [.andersen blackstad @307@, bliss gardner-medwin @1023@, bliss lomo @1022@.] , but it may also be observed in the two remaining components of the trisynaptic pathway, the mossy fiber pathway and the Schaffer-collateral pathway [.alger teyler slice 1976 @2125@.].  It is most commonly induced by high-frequency stimulation leading to a brief induction phase during and following the stimulation, and a maintenance phase in which the potentiation of synaptic transmission lasts up to several hours.  It is measured experimentally by stimulating a group of presynaptic neurons and measuring the excitatory postsynaptic potential (EPSP) of the postsynaptic cells.  After a period of high-frequency stimulation, the original low-frequency stimulation is resumed, at which time there is a persistent increase in the size of the EPSP.  These long-lasting increases in synaptic efficacy make LTP a popular candidate mechanism for the storage of information as well as for experience-dependent development.

	To account for development and plasticity throughout the lifetime of an organism, however, there must be a mechanism by which the connections between neurons can be weakened or eliminated in an activity-dependent fashion.  Such a mechanism appears to exist in the form of long-term depression.  Like LTP, LTD has been observed in the hippocampus as well as in other brain regions.  While LTP appears to require synchronous pairing of synaptic activation and postsynaptic depolarization, LTD appears to be induced by asynchronous pairing of such pulses.  Homosynaptic LTD, the form of LTD that involves depression of a pathway as a result of activity of that same pathway, is characterized by a lasting decrease in synaptic strength, and is usually induced by prolonged application of stimulation at frequencies below those required to induce LTP.  Common paradigms for the induction of homosynaptic  LTD include the application of low-frequency stimulation, such as 1 Hz or 3 Hz for 5--15 min [.dunwiddie lynch depression 1978 @3152@, dudek bidirectional 1993 @11444@.].  As the tetanus is applied, the amplitude of the EPSP decreases gradually until it reaches an amplitude as much as 35\% smaller than control, and persists at this level for more than one hour.

\section{Postsynaptic Ca$^{2+}$ and synaptic plasticity}

\subsection{The Lisman hypothesis}

	Studies indicate that postsynaptic Ca$^{2+}$ is a common feature to LTD and LTP and that LTD- and LTP-inducing stimuli result in increases in postsynaptic [Ca$^{2+}$] [.Christie learning 1996 @13825@, Regehr Tank 1990 @9999@.].  Application of Ca$^{2+}$ chelators has been shown to block induction of LTD and LTP [.Malenka Lancaster 1992 @11059@, Bolshakov Siegelbaum 1994 @11961@.], and reductions in extracellular Ca$^{2+}$ concentrations alter the probability of LTD and LTP induction [.Christofi 1993 @11416@, Artola Hensch 1996 @13710@.].  Thus,  it is of interest how such a shared mechanism can give rise to the opposing processes of LTD and LTP.  In a hypothesis put forth in 1989 by John Lisman, it was suggested that the magnitude of the Ca$^{2+}$ increase determines the nature of the effect, with small increases contributing to LTD and large increases contributing to LTP.  In general, low-frequency stimulation results in LTD [.Dudek Bear 1992 @10909@.] while higher frequency stimulation results in LTP [.Kirkwood science 1993 @11560@,  Teyler Little 1994 @12482@.].  It is likely that low-frequency stimulation leads to small, sustained increases in [Ca$^{2+}$]$_{i}$, while high-frequency stimulation leads to larger increases in [Ca$^{2+}$]$_{i}$. Small increases in postsynaptic [Ca$^{2+}$] are thought to activate phosphatases, leading to LTD induction, whereas larger increases in postsynaptic [Ca$^{2+}$] are thought to activate kinases, leading to LTP induction [.Lisman 1989 @9371@.]. 

	This differential activation of kinases or phosphatases is thought to occur because kinases and phosphatases have different affinities for Ca$^{2+}$/calmodulin (Ca$^{2+}$/CaM), phosphatases having high affinity and kinases having low affinity.  For example, when small increases in intracellular Ca$^{2+}$ occur, Ca$^{2+}$/CaM activates the phosphatase calcineurin which has a high affinity for it.  Calcineurin in turn dephosphorylates and inactivates Inhibitor 1 (I1) disinhibiting protein phosphatase 1 (PP1).  PP1 then dephosphorylates various substrates, including many cellular kinases, and LTD is obtained.  Conversely, when large increases in intracellular Ca$^{2+}$ occur, a greater concentration of Ca$^{2+}$/CaM is available to activate the low affinity kinases, resulting in the phosphorylation of various substrates, including kinase autophosphorylation, and LTP is obtained.  	 

\subsection{Ca$^{2+}$ channels in LTP and LTD}

	Several routes of Ca$^{2+}$ entry have been implicated in LTP and LTD.  The NMDA receptor is a voltage-dependent, ligand-gated channel that requires both glutamate binding and sufficient depolarization to remove a Mg$^{2+}$ block from the receptor.   NMDA receptor-mediated Ca$^{2+}$ influx has been shown to be involved in both LTD [.Dudek Bear 1992 @10909@, Selig NMDA 1995 @12979@, Mulkey Malenka 1992 @11104@.] and LTP [.Bliss Collingridge 1993 @11376@, Madison Rev 1991 @10449@.].  

	Studies have shown that activation of afferent fibers in the hippocampal stratum radiatum such as that applied during induction of LTP and LTD gives rise to a large influx of Ca$^{2+}$that is predominantly due to the activation of voltage-gated Ca$^{2+}$ channels (VGCCs) [.miyakawa neuron 1992 @11090@,  magee science 1995 @12705@.]. CA1 pyramidal cells of the hippocampus express at least five  different VGCCs including low threshold T channels and high threshold L, N, R, and P channels [.fisher properties 1990 @9775@, mintz adams 1992 @11086@ .], and fluorescence imaging studies performed in this laboratory have shown these channels to have a characteristic distribution.  High-threshold channels such as N-, P-, Q- and L-type channels are preferentially located in the somatic region, while low-threshold channels such as T-type and the high-threshold R-type are localized to the distal dendrites [.christie eliot 1995 @12874@.].  Other recent data from this laboratory indicate the existence of a dihydropyridine-sensitive channel (presumably L-type) active at resting membrane potential and which is located at the soma and throughout most of the dendritic tree [.magee avery 1996 @13973@.]. VGCC-mediated Ca$^{2+}$ influx through several of the channel subtypes has been shown to be involved in the induction of both LTD [.Bolshakov Siegelbaum 1994 @11961@, Christie Schexnayder 1997 @14599@, Christie Learning 1996 @13825@, Wang Rowan LTD 1997 @14479@.] and LTP [.Magee Science 1997 @14316@, Grover Teyler Nature 1990 @9815@, Ito Miura 1995, Wang Rowan LTP 1997 @15621@.] .  

\subsection{Sliding threshold models}

	The sliding modification threshold rule or BCM rule (for Bienenstock, Cooper, and Munro) was originally used to model development of visual cortex.  It stated that a previous history of a high level of synaptic activity would shift the modification threshold ($\Theta_{M}$) such that increased strength of synaptic connections would become less likely and decreases in strength would become more likely [.bienenstock cooper @4773@.].  In this way, neurons would be better able to maintain connection strength in the linear range and to maximize information storage.  This rule has been applied to the problem of LTD/LTP induction, and it has been suggested that the threshold for LTD vs LTP ($\Theta_{LTD/P}$)  "slides" or varies as a function of the average activity of the postsynaptic neuron, such that low frequencies of stimulation result in LTD while higher frequencies result in LTP.  

	As postsynaptic Ca$^{2+}$ is known to play a role in initiation of LTD and LTP, and is known to be related to the frequency of stimulation, it is possible that the threshold for LTD vs LTP may be related to postsynaptic Ca$^{2+}$.  This hypothesis is best described by the ABS rule (Artola, Brocher, and Singer) for activity-dependent synaptic modifications.  It states that the direction of the synaptic gain change depends on the membrane potential of the postsynaptic cell, or more mechanistically, on the amplitude of the surge of intracellular Ca$^{2+}$ concentration [.artola singer trends @11345@.].  This makes it possible to relate the two voltage-dependent thresholds for LTD and LTP to corresponding changes in [Ca$^{2+}$]$_{i}$, as the influx of Ca$^{2+}$ through NMDA receptors and voltage-gated Ca$^{2+}$ channels increases with depolarization.  Hence, the main determinant for the polarity of changes in synaptic strength would be the amplitude of the Ca$^{2+}$ increase, rather than the source of the Ca$^{2+}$.  This hypothesis fits well with the Lisman hypothesis regarding the importance of the magnitude of increases in postsynaptic Ca$^{2+}$ and the induction of LTD or LTP.  Indeed, others have shown that when increases in postsynaptic [Ca$^{2+}$] are reduced by blockade of NMDA receptors [.Artola Nature 1990 @9647@, Cummings Neuron 1996 @13586@.], reductions in extracellular  [Ca$^{2+}$] [.Mulkey Malenka 1992 @11104@.], or buffering of intracellular [Ca$^{2+}$] [.Kimura 1990, Brocher chelators 1992, Hansel 1997 @15913@.], stimulation protocols that usually result in LTP give rise instead to LTD.	 

\section{Dendritic K$^{+}$ channels and long-term potentiation}

\subsection{Transient A-type K$^{+}$ channels in CA1 dendrites}

The membrane excitability of neurons is believed to be regulated primarily by voltage-dependent K$^{+}$ channels which act by controlling membrane potential and the threshold for generation of action potentials as well as  shaping excitatory postsynaptic potentials (EPSPs).  Recent studies have shown the transient or A-type K$^{+}$ channel to be distributed throughout the dendrites of hippocampal CA1 pyramidal cells, increasing in density with distance from the soma [.hoffman nature @15041@.].  This makes the transient K$^{+}$ channel an excellent candidate for controlling signal propagation in the dendrites.  It is likely that this transient K$^{+}$ channel corresponds to the Kv4.2 type of K$^{+}$ channel which has been characterized in the hippocampus by immunohistochemical techniques [.sheng 1992 @14549@.].   

\subsection{Kinase modulation of dendritic K$^{+}$ channels}

The amino acid sequence of the Kv4.2 K$^{+}$ channel has been shown to contain phosphorylation sites for protein kinases [.baldwin 1991 @14544@, adams sweatt biol @16738@.]. Activators of cyclic-AMP dependent protein kinase (PKA) and protein kinase C (PKC) have been shown to increase the amplitude of EPSPs and population spikes in hippocampal neurons [.slack efficacy @14547@,hu albert @15164@.]. In addition, activators of PKA and PKC appear to shift the activation curves of these dendritic K$^{+}$ channels to more positive potentials and to increase the amplitude of back-propagating action potentials [.hoffman downregulation @15975@.]. Thus, it appears that phosphorylation of these K$^{+}$ channels can result in downregulation of their activity, providing an attractive mechanism for increases in neuronal excitability.             

\subsection{Implications for A-channel modulation in LTP induction}

Increased kinase activity has been well documented in the induction of LTP, though the substrates for this increased phosphorylation and the means by which potentiation occurs have yet to be well characterized. It is possible that these dendritic K$^{+}$ channels are an important substrate for the kinases involved in LTP and that the persistent downregulation of dendritic transient K$^{+}$ channel activity described above may contribute to lasting increases in neuronal excitability, such as those observed in LTP.

\chapter*{2 Methods and Materials}
\setcounter{chapter}{2} \addcontentsline{toc}{chapter}{2 Methods and
Materials}

\setcounter{section}{0} \setcounter{subsection}{0}

\section{Slice preparation}

	Hippocampal slices (400~$\mu$m) were prepared from young Sprague-Dawley Rats (14--21~days or 6-7 weeks).  Brains were rapidly dissected in cold saline and slices were prepared with a Vibratome.  Slices were immediately placed in a submerged, oxygenated holding chamber where they were incubated at 35$\pm$1$^\circ$C for at least 20 min, and thereafter at room temperature (22$\pm$1$^\circ$C).  External solution consisted of (in mM) 124 NaCl, 26, NaHCO$_3$, 10 Dextrose, 2.5 KCl, 2.5 CaCl$_2$, 1.5 MgCl$_2$, and 1.25 NaH$_2$PO$_4$ and was bubbled with 95\% O$_2$-5\% CO$_2$.  Individual slices were transferred, as needed, to a submerged recording chamber held at 32$\pm$1$^\circ$C unless otherwise noted.

\section{Extracellular recordings}

Extracellular recordings were carried out at either 23 or 32$^\circ$C using conventional field recording methods in a submerged chamber.  Field EPSPs (f-EPSPs) were recorded from the stratum radiatum in the CA1 region using glass electrodes filled with 1 M NaCl.  Constant current stimuli were delivered with a bipolar tungsten wire electrode placed in stratum radiatum immediately adjacent to the recording electrode.  To monitor the f-EPSP slope, individual stimuli were were administered every 15 s.  To induce LTD, low-frequency stimulation (LFS; 900 pulses) was administered at either 1 or 3 Hz using the same intensity stimuli used to monitor f-EPSPs. D,L-2-amino-5-phosphonovaleric acid(D,L-APV; 50~$\mu$M) and NiCl$_{2}$ (25~$\mu$M) were prepared daily from 5 mM stock solutions.  Nimodipine (10~$\mu$M) was prepared daily in ethanol, and all recordings were performed in a darkened room.  Data were expressed as means $\pm$SEM of 20 responses recorded over a 5 min period (25-30 min postconditioning) unless otherwise stated.  Data points in figures represent averages of 20 responses recorded over 5 min periods.  Statistical analysis was performed using unpaired t-tests, with significance levels set at P=0.05.   

\section{Whole-cell recordings}

 All whole-cell recordings were carried out at 32$^\circ$C. A Zeiss Axioskop, fitted with a 40x water immersion objective and differential interference contrast optics (DIC), was used to view the slices.  Light in the infrared range (740~nm) was used to better resolve individual neurons and their dendrites.  Whole-cell recording electrodes (2--4 megohm) were pulled from borosilicate glass (Drummond) and filled with a solution consisting of (in mM) 120 K gluconate, 20 KCl, 10 HEPES, 4 NaCl, 4 Mg-ATP, 0.3 Tris-GTP, 14 phosphocreatine, and 0.08 fura-2 (pH 7.25 with KOH).  Whole-cell recordings were made from visually identified CA1 pyramidal neurons located within 50~$\mu$m of the surface of the slice using an Axoclamp 2A amplifier in bridge mode (Axon Instruments).  Neurons exhibited a resting membrane potential of --60 to --70 mV.  Series resistance for these recordings was 10 to 20 megohms.  A single, etched tungsten wire (<5~$\mu$m diameter) was placed near the dendrite of the neuron of interest ($\sim$100~$\mu$m from the soma) for focal extracellular stimulation.  The pairing protocol used to induce LTD/LTP  in these experiments consisted of small amplitude ($\sim$5~mV) excitatory postsynaptic potentials (EPSPs) elicited by synaptic stimulation paired with back-propagating action potentials evoked by depolarizing current pulses in the soma (2~nA for 2~ms).  The small amplitude EPSPs were utilized for the test pulses to ensure that the EPSPs would remain subthreshold following the induction of LTP without having to change stimulus strength.  In experiments where drugs were added to the external saline, nimodipine (10~$\mu$M) was prepared daily from 10 mM stock solution (in ethanol) and D,L-APV (10 or 50~$\mu$M) was prepared daily from 25 mM stock solution.

\begin{figure}[p]
\centerline {\epsfxsize=5in 
\epsffile{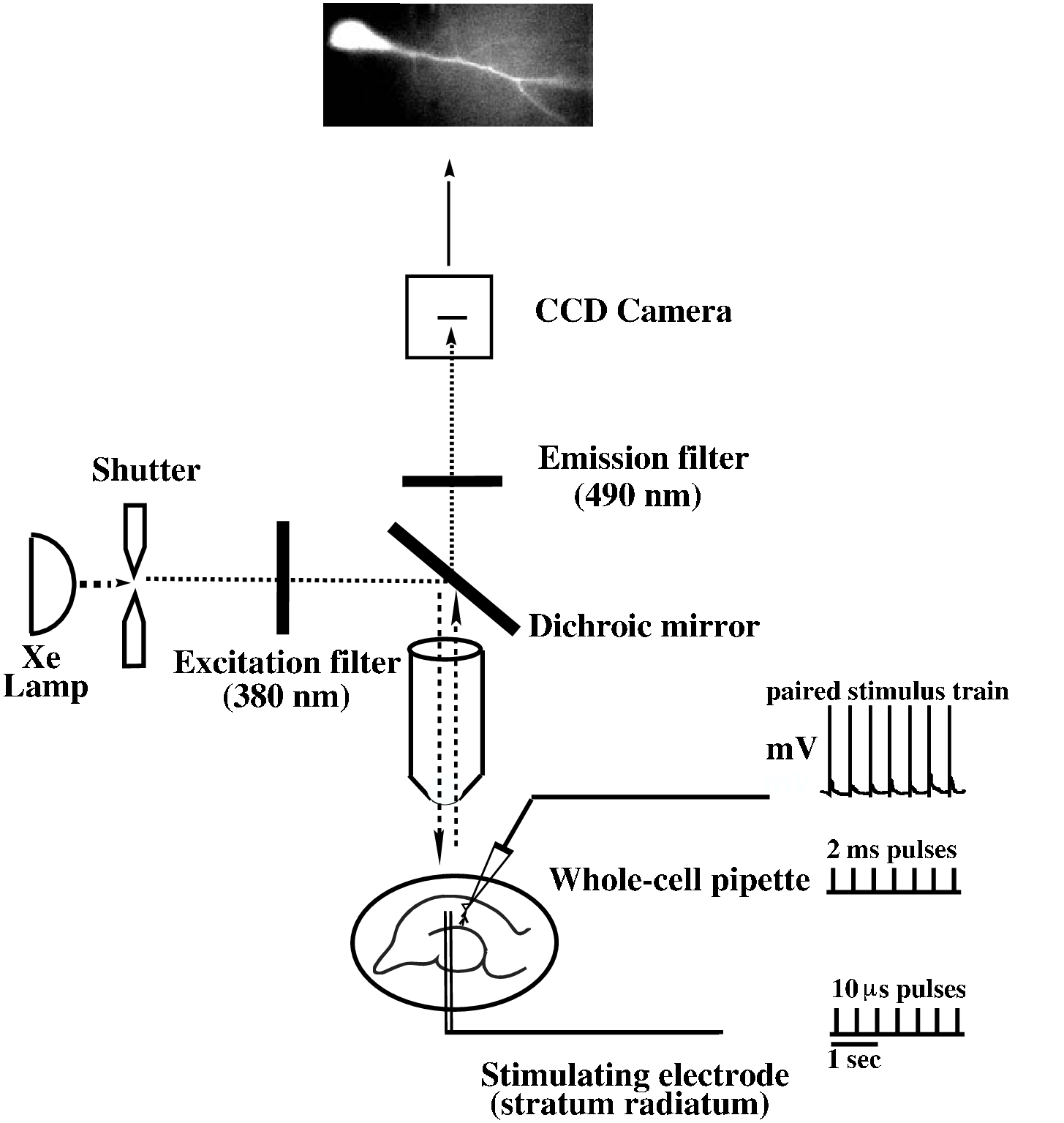}}

\caption[Schematic diagram of equipment used for electrical recording and fluorescence imaging.]
{{\bf Schematic diagram of equipment used for electrical recording and fluorescence imaging.} {Whole-cell recordings were obtained from the soma of an individual neuron. Pairing stimulation consisted of excitatory postsynaptic potentials (EPSPs) elicited by synaptic stimulation with an extracellular electrode placed in stratum radiatum paired with back-propagating action potentials evoked by the application of depolarizing current pulses to the soma through the whole-cell electrode. The timing of the EPSP and current pulse produced action potentials that occurred at the approximate peak of the EPSP.  Fluorescence data were recorded with a cooled CCD camera.}}
\label{Fig1}
\end{figure}

\section{Ca$^{2+}$ imaging}

Imaging of [Ca$^{2+}$]$_{i}$ was performed concurrently with the above electrical recordings in each neuron. To measure changes in [Ca$^{2+}$]$_{i}$, the fluorescent indicator fura-2 (80~$\mu$M, K$_{d}$=230 nM) was included in the pipette solution, and each cell was allowed to dialyze for 15 min before optical recordings were obtained.  A cooled, charge-coupled device (CCD) camera (Photometrics, Tucson, AZ) in sequential frame-transfer mode [.lasser-ross young @10422@.]  was used to record high-speed fluorescence images of a 200~$\mu$m length of soma and dendrite (Fig. 1) [.christie learning 1996 @13825@.].  Relative changes in [Ca$^{2+}$]$_{i}$ were quantified as changes in $\Delta$F/F (Fig. 2), where F is fluorescence intensity before stimulation (after subtracting autofluorescence) and $\Delta$F is the change from this value during neuronal activity (corrected for bleaching during the run).  The bleaching correction was obtained by measuring the fluorescence of the neuron under nonstimulated conditions.  Tissue autofluorescence was obtained at the end of each experiment by making an equivalent measurement from a parallel location in the slice away from the dye-filled neuron.  Light at 380~nm (13 nm bandpass filter, Omega Optical) was used to excite the fura-2.  A 490 nm emission filter was used.  Recordings were made with 20 ms frame intervals, and pixels were binned in a 10 by 10 array.  Using the 40x objective it was possible to image the soma and the first 150-200~$\mu$m of the proximal apical dendrite.  Within this field of interest, three locations were used for analysis including the soma, the first 25~$\mu$m of the proximal dendrite (proximal), and an intermediate region of the dendrite about 100~$\mu$m from the soma (midradiatum).  All fluorescence measurements were carried out in a darkened room. It should be noted that to measure changes in dendritic Ca$^{2+}$during trains of back-propagating action potentials or subthreshold EPSPs, imaging was performed as above, but with the use of the fluorescent Ca$^{2+}$ indicator BIS-fura (110 ~$\mu$M).

\begin{figure}[p]
\centerline {\epsfxsize=4in 
\epsffile{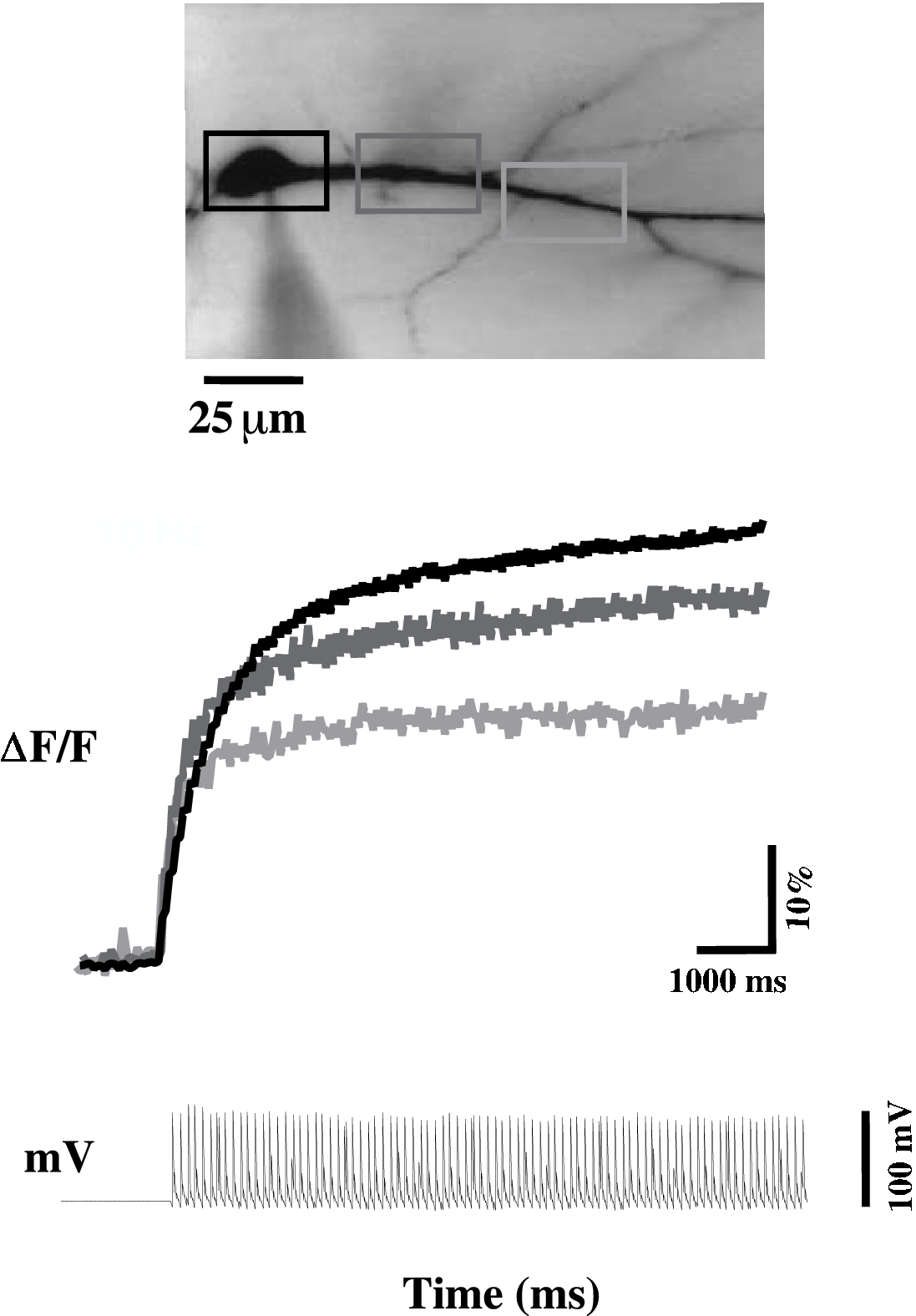}}

\caption[Sample fluorescence and electrical traces during 10 Hz stimulation.]
{{\bf Sample fluorescence and electrical traces during 10 Hz stimulation.} {Fluorescence traces (\%$\Delta$F/F) represent the changes in [Ca$^{2+}$] for three regions of the fura-2-filled neuron (above) in response to a 10 s stimulus train at 10 Hz.  Bottom trace is the electrical response measured in the soma from the paired EPSPs and action potentials.}}
\label{Fig2}
\end{figure}

\section{Stimulus protocols and analysis}

Stable electrical recordings consisting of EPSPs of $\sim$5 mV amplitude were obtained for at least 10 min before LTD/LTP induction protocols were delivered.  In experiments where blockers were used, an additional 10 min of baseline recordings were obtained after nimodipine or APV were added to the bath unless otherwise noted.  Input resistance was monitored throughout each experiment.  The stimuli used to induce LTD/LTP were delivered at 3, 10, 30, 50, 100, or 200~Hz.  For the 3, 10, 30, and 50~Hz stimulation protocols, 900 pulses were applied in 9 epochs of 100 pulses each.  For the 100 and 200~Hz stimulation protocols, 400 pulses were applied in 4 epochs of 100 pulses each.  All stimuli were applied with intervals of 5 s between epochs.  For each induction procedure the reported changes in $\Delta$F/F are based on measurements of the peak $\Delta$F/F obtained near the end of the first epoch of 100 pulses (or after 10 s for 3 Hz stimulation), averaged over several experiments, and are expressed as mean $\pm$SEM .  Data for percent LTD or LTP are expressed as mean percent change in EPSP slope ($\pm$SEM), averaged over 5 min of responses obtained 25--30 min after the induction protocol, and averaged over several experiments.  

In experiments where changes in dendritic Ca$^{2+}$ were measured during trains of back-propagating action potentials, LTP was induced with 100 Hz stimulation consisting of 400 pulses applied in 4 epochs of 100 pulses each, separated by 5s.  The extracellular stimulating electrode was placed 150-200 $\mu$m from the soma.  Trains of action potentials used to evoke dendritic Ca$^{2+}$signals consisted of 5 action potentials at 20 Hz.  Changes in dendritic Ca$^{2+}$induced by these trains were measured at the peak of the Ca$^{2+}$ signal and averaged over 15 traces, and are expressed as mean change in amplitude of Ca$^{2+}$ signal $\pm$SEM for each region of the dendrite.  Significance was determined with paired, two-tailed t-tests.  Trains of subthreshold EPSPs used to determine the location of the synapse of interest consisted of 5 EPSPs of 20-30 mV at 50 Hz.  These trains were delivered 400 ms after a 1500 ms hyperpolarizing prepulse of 15-25 mV.  It should be noted that fluorescence imaging of Ca$^{2+}$ was also performed during the tetani used to induce LTP, but these data were not analyzed.       

\chapter*{3 Results}
\setcounter{chapter}{3}
\addcontentsline{toc}{chapter}{3 Results}
\setcounter{section}{0}

\section{Postsynaptic Ca$^{2+}$ in Homosynaptic LTD Induction}

\subsection{Temperature dependence}

It has been shown that, when slices are maintained at higher temperatures, LTP of a greater magnitude can be induced than with the same stimulus protocol at room temperature [.Williams Nayak 1993 @11874@.].  To test the hypothesis that low-frequency-induced homosynaptic LTD might also exhibit a temperature dependence, trains of conditioning stimuli were administered at either room temperature (23$^\circ$C) or at 32-35$^\circ$C, and f-EPSPs were recorded [.schexnayder depression @14599@.].  A significant depression of synaptic efficacy was observed at room temperature, for both the 1 Hz stimuli (-20.17 $\pm$7.5\%, n=6, P<0.05; Fig. 3A) and 3 Hz stimuli (-22.17 $\pm$2.2, n=5, P<0.05; Fig. 3B).  In contrast to what has been observed for LTP in this region, increasing the bath temperature to 32-35$^\circ$C did not produce significantly greater LTD with either frequency (1 Hz: -20.72 $\pm$7.5, n=6; 3 Hz: -25.5 $\pm$10.2, n=5; Fig 3, A and B).  This indicates that homosynaptic LTD induction is not as temperature-sensitive as LTP.  To ensure that the LTD under investigation was similar to that observed by others [.bolshakov science 1994 @11961@, dudek bear 1992 @10909@, mulkey malenka 1992 @11104@,  selig bear 1995 @13073@.], 50~$\mu$M D,L-APV was added to the bath in an attempt to block the induction of the LTD.  As shown in Fig 3C, D,L-APV completely blocked the induction of homosynaptic LTD with both the 1 Hz and 3 Hz stimulus paradigms (1 Hz: 5.0 $\pm$13.2, n=6; 3 Hz: 27.5 $\pm$9.8, n=8; Fig. 2C), although LTD could still be induced after washout of the drug (1 Hz: -36.8 $\pm$12.4; 3 Hz: -18.0 $\pm$6.5; Fig. 3C).

\begin{figure}[p]
\centerline {\epsfxsize=3.5in 
\epsffile{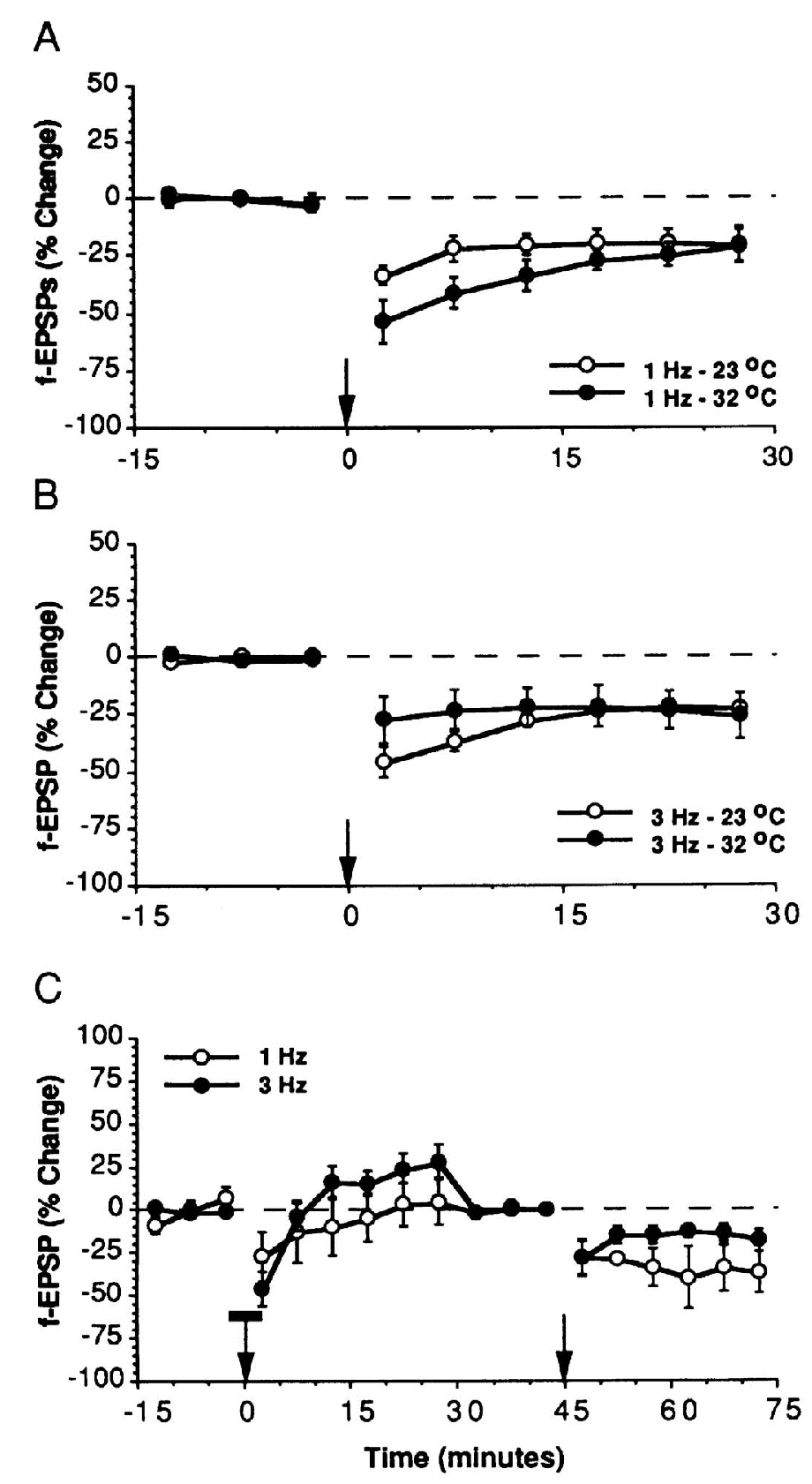}}

\caption[Homosynaptic long-term depression (LTD) induction is not temperature dependent.]
{{\bf Homosynaptic long-term depression (LTD) induction is not temperature dependent.} {{\bf (A)\/}: Application of 900 pulses at 1 Hz produced a significant depression of evoked field responses at both 23 and 32$^\circ$C.  No difference in the magnitude of LTD was evident at 30 min postconditioning.  {\bf (B)\/}:  3 Hz stimuli also produced significant and equivalent LTD at 23 and 32$^\circ$C.  {\bf (C)\/}:  Application of low-frequency stimulation (LFS) in the presence of APV (solid bar) does not result in the induction of homosynaptic LTD with either the 1 or 3 Hz stimuli.  When the same stimuli are applied in control conditions LTD is produced.  Arrows represent the time point at which conditioning stimuli were applied.}}
\label{Fig3}
\end{figure}

\subsection{Effects of external Ca$^{2+}$ levels}

To examine the involvement of Ca$^{2+}$, we initially tested field responses in slices perfused with ACSF containing either low (Ca$^{2+}$-free), normal (2.5 mM), or high (4 mM) Ca$^{2+}$.  For the low-Ca$^{2+}$ condition, slices were first tested in normal ACSF to acquire stable baseline recordings.  The normal ACSF was then switched to one that contained 0 mM Ca$^{2+}$, and perfused for 5 min to reduce the bath Ca$^{2+}$ level significantly.  During this period, responses diminished to 0--5\% of their initial size, and then 3 Hz stimulation was applied while Ca$^{2+}$-free perfusion continued for an additional 5 min.  After the application of the 3 Hz stimuli, the solution was switched back to one containing normal Ca$^{2+}$.  Slices bathed in a Ca$^{2+}$ free solution immediately before and during the low-frequency trains did not exhibit homosynaptic LTD, as measured 30 min postconditioning (9.04 $\pm$5.0, n=4, P>0.05; Fig. 4A).  Slices that were then given the low-frequency stimulation again in normal Ca$^{2+}$ exhibited LTD of the field response similar to that shown previously (-28.8 $\pm$10.1, n=4, see also Fig. 3B).  In the elevated Ca$^{2+}$ condition, recordings were made from slices continuously bathed in ACSF containing 4 mM Ca$^{2+}$ following the acquisition of baseline responses in normal ACSF.  In these animals, even though the size of the fiber volley remained unchanged, the amplitude and the slope of the f-EPSP increased in the presence of the 4 mM Ca$^{2+}$.  In this condition the application of the 3 Hz stimuli also failed to produce homosynaptic LTD and, in fact, produced a slight potentiation (16.7 $\pm$12.0, n=7; Fig 4B).  When the external Ca$^{2+}$ was then lowered back to 2.5 mM, it was possible to depress the evoked responses from their elevated levels to a point below the original baseline level (-10.0 $\pm$18.8, n=7). 

\begin{figure}[p]
\centerline {\epsfxsize=4in 
\epsffile{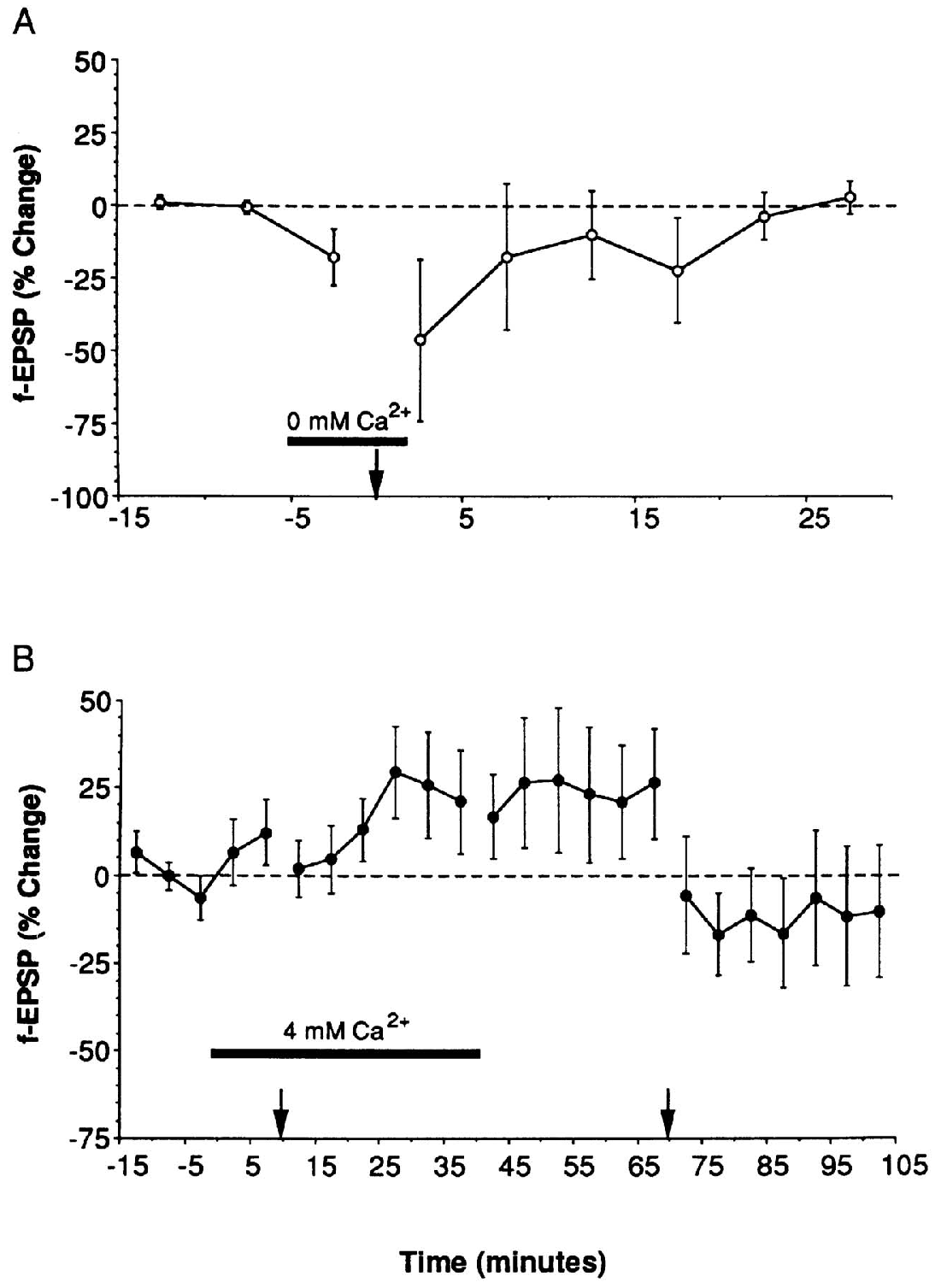}}

\caption[External Ca$^{2+}$ levels influence the expression of homosynaptic LTD.]
{{\bf External Ca$^{2+}$ levels influence the expression of homosynaptic LTD.} {{\bf (A)\/}: Field responses recorded before and after perfusion of the slice with a Ca$^{2+}$-free solution.  LFS (3 Hz) was administered in the presence of the 0 mM Ca$^{2+}$ bathing medium.  Responses recorded after the restoration of normal external Ca$^{2+}$ levels fail to exhibit LTD.  {\bf (B)\/}:  Increasing extracellular Ca$^{2+}$ to 4 mM (solid bar) disrupts the induction of LTD and a slight potentiation is produced by the LFS.  When the extracellular medium is switched back to one containing 2.5 mM Ca$^{2+}$, application of the LFS depresses evoked reponses.  Arrows represent time points at which conditioning stimuli were applied.}}
\label{Fig4}
\end{figure}

\subsection{Contribution of voltage-gated Ca$^{2+}$ channels}

Patch-clamp and fluorescence imaging data indicate that the L-, R-, and T-type VGCCs are the primary VGCCs active in the apical dendrites of CA1 pyramidal cells [.christie eliot 1995 @12874@, magee science 1995 @12705@, magee christie 1995 @13563@.].  Previously, there has been some controversy regarding the sensitivity of homosynaptic LTD to L-type VGCC antagonists [.bolshakov science 1994 @11961@, selig bear 1995 @13073@.].  LTD was blocked by the L-type antagonist nitrendipine [.bolshakov science 1994 @11961@.], whereas no effect was observed when nifedipine was used to block L-type channels in young animals [.selig bear 1995 @13073@.].  For our experiments we chose to use nimodipine, a potent and relatively nonphotolabile blocker of L-type VGCCs that has been used to block heterosynaptic LTD both in vitro (Wickens and Abraham, 1991) and in vivo, [.Christie abraham l-type 1994 @12000@.].  Because nimodipine may cause a small increase in evoked EPSPs when bath applied, this compound was included in the bath continuously for these experiments [.o'regan 1991 @10521@.].  Nimodipine (10~$\mu$M) prevented the induction of homosynaptic LTD with both the 1 Hz and 3 Hz stimulation (1 Hz: 0.8 $\pm$9.4, n=4; 3 Hz: 12.3 $\pm$8.4, n=4; Fig 5).  Due the the lack of a specific blocker for the R- and T-type VGCCs, we used NiCl$_{2}$, which acts on both of these channels [.avery ca3 1996 @13792@, magee science 1995 @12705@.].  All experiments were performed using 25~$\mu$M NiCl$_{2}$, a concentration that we (Schexnayder et al., 1995) and others [.ito miura 1995 @13613@.] have found does not alter synaptic transmission.  When 25~$\mu$M NiCl$_{2}$ was applied to the slice during the induction procedure, LTD induction was prevented with both the 1 and 3 Hz stimulation protocols (1 Hz: 2.5 $\pm$11.2, n=6; 3 Hz: 12.6 $\pm$12.1, n=5; Fig. 5B).  Attempts to induce LTD 30-45 min after NiCl$_{2}$ washout invariably failed, despite the fact the f-EPSPs appeared unaffected by the drug.

\begin{figure}[p]
\centerline {\epsfxsize=4in 
\epsffile{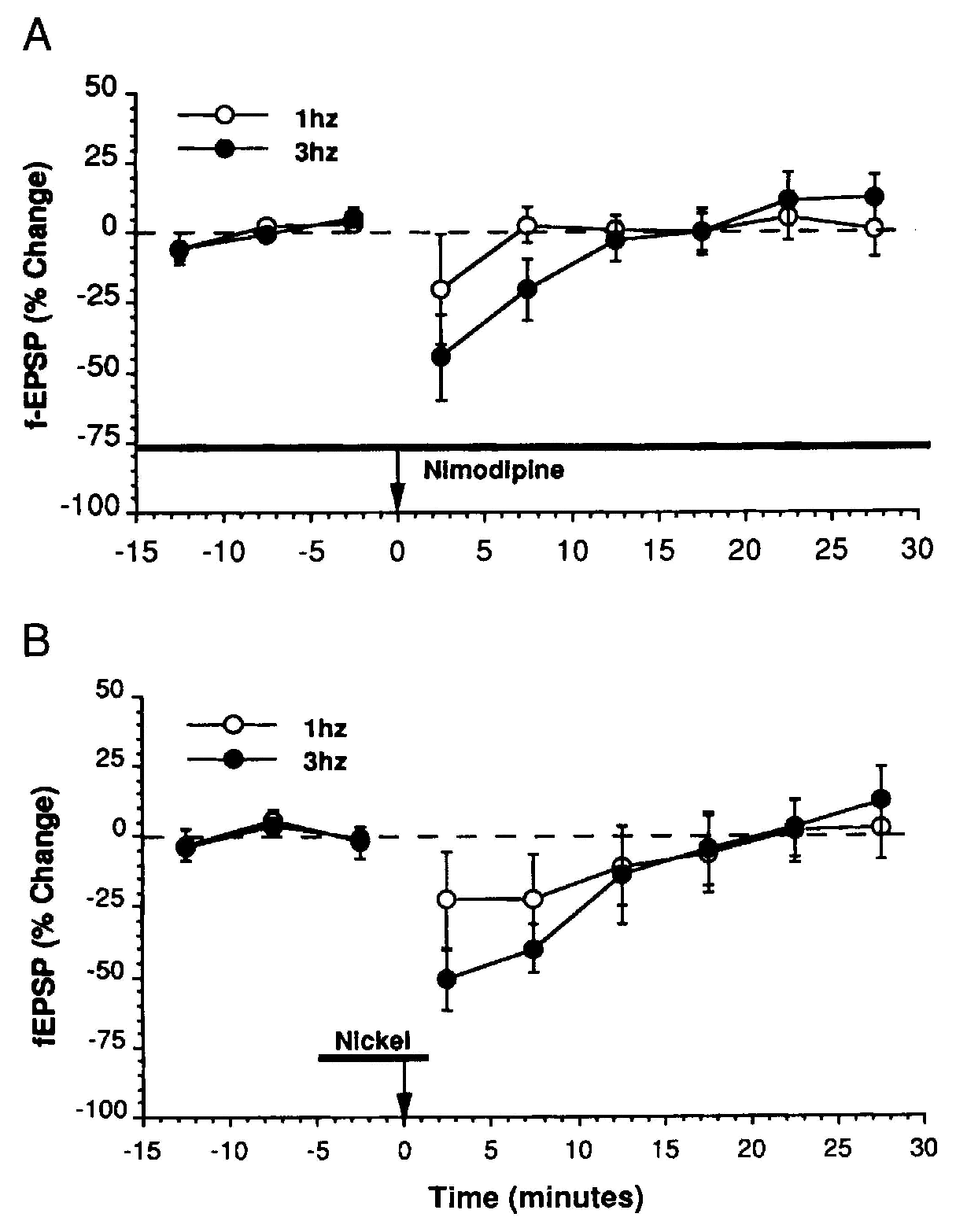}}

\caption[Blockade of voltage-gated Ca$^{2+}$ channels (VGCCs) prevents LTD induction.]
{{\bf Blockade of voltage-gated Ca$^{2+}$ channels (VGCCs) prevents LTD induction.} {{\bf (A)\/}:  L-type VGCC antagonist nimodipine (10 $\mu$M) prevented the induction of LTD with both the 1 and 3 Hz stimuli.  {\bf (B)\/}:  Addition of the R- and T-type channel antagonist NiCl$_{2}$ (25 $\mu$M) to the recording medium during the application of the LFS also prevents the induction of homosynaptic LTD.  Arrows represent time points at which conditioning stimuli were applied.}}
\label{Fig5}
\end{figure}

\section{ Postsynaptic Ca$^{2+}$ Imaging during LTD/LTP Induction}

\subsection{Measurements of postsynaptic [Ca$^{2+}$] and correlation with frequency of stimulation and magnitude and direction of changes in synaptic strength}

	Fluorescence imaging experiments demonstrated that the level of postsynaptic Ca$^{2+}$ varied with frequency of stimulation, as well as with the magnitude and direction of change in synaptic strength (Fig. 6).  The Ca$^{2+}$ data are expressed as changes in fluorescence  (\%$\Delta$F/F),  measured in three regions of the neuron, including the soma and proximal and middle dendrites.  LTD was achieved with stimulation from 3 to 10 Hz while LTP was achieved with stimulation of 30 Hz and above.  Although there was no precise magnitude of postsynaptic Ca$^{2+}$ signal that corresponded to the division between LTP and LTD, the shaded area represented a range of values above which LTP was observed and below which LTD was observed.

\begin{figure}[p]
\centerline {\epsfxsize=6in 
\epsffile{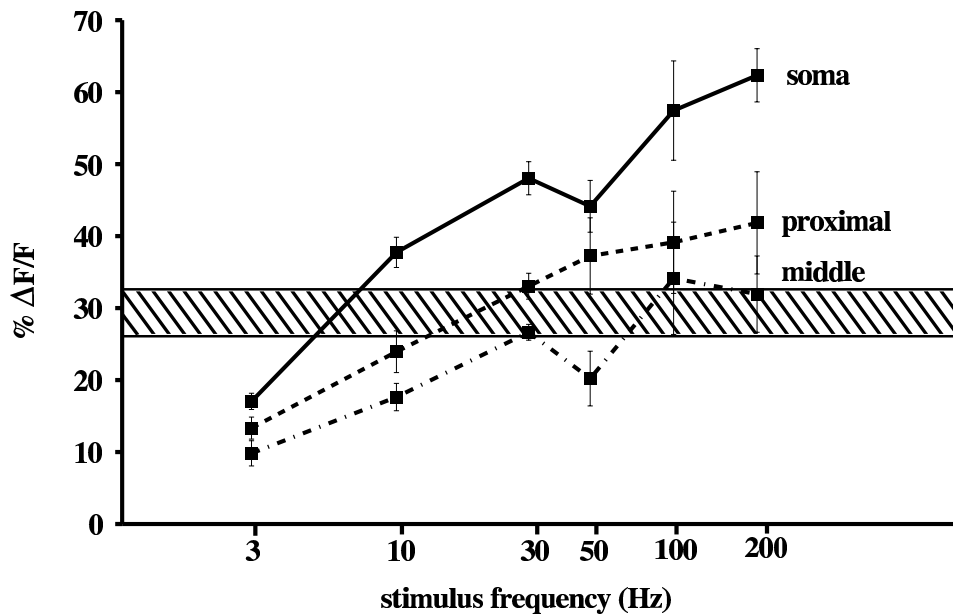}}

\caption[Postsynaptic Ca$^{2+}$ varies with stimulus frequency.]
{{\bf Postsynaptic Ca$^{2+}$ varies with stimulus frequency.} {The shaded area represents a range of postsynaptic [Ca$^{2+}$] values below which LTD was observed (3 and 10 Hz stimulation) and above which LTP was observed (30 Hz and above).  The [Ca$^{2+}$]$_{i}$ data are expressed as changes in fluorescence (\% $\Delta$F/F), and include values for three regions of the neuron: the soma, proximal dendrite, and middle dendrite. (n = 3 -- 7 for each point)}}
\label{Fig6}
\end{figure}

\subsection{ Reductions in stimulus-induced [Ca$^{2+}$]$_{i}$ in the presence of nimodipine or APV}

Stimuli of various frequencies were applied in the presence or absence of 10~$\mu$M nimodipine, or in the presence or absence of 10 or 50~$\mu$M APV (Fig. 7).  In each region Ca$^{2+}$ signals increased with increasing stimulus frequency.   In the soma nimodipine appeared to block Ca$^{2+}$signals by about 30\% at each stimulus frequency, but 10 or 50~$\mu$M APV appeared to have little effect on Ca$^{2+}$signals in  the soma.  In the proximal and middle dendritic regions, however, both nimodipine and 10~$\mu$M APV appeared to block Ca$^{2+}$ signals to a similar extent at most of the frequencies tested.  The block by 50~$\mu$M APV was slightly greater than by 10~$\mu$M APV.

\begin{figure}[p]
\centerline {\epsfxsize=4.5in 
\epsffile{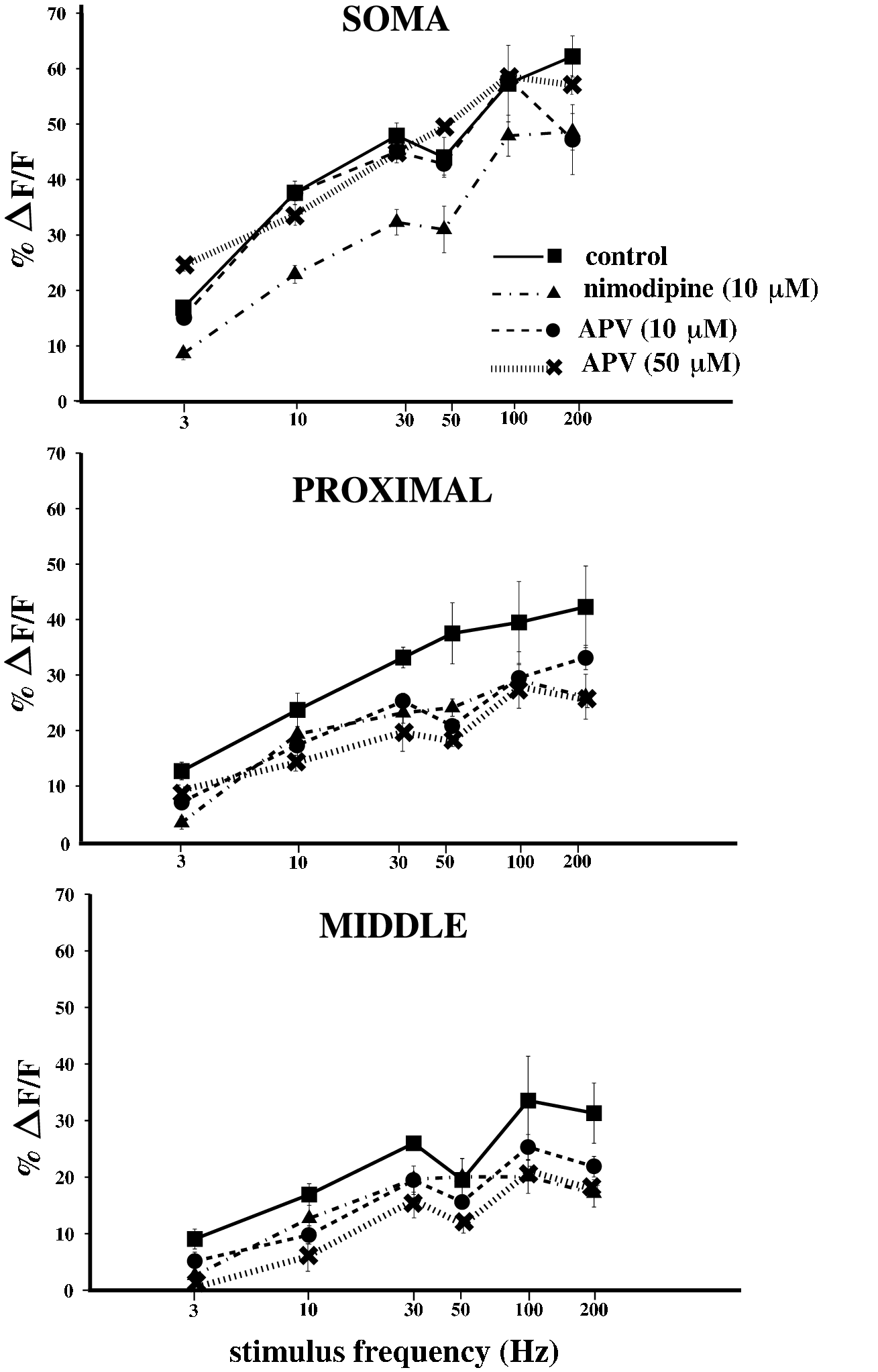}}

\caption[Application of 10 $\mu$M APV, 50 $\mu$M APV, or 10 $\mu$M nimodipine reduced postsynaptic Ca$^{2+}$ induced by different frequencies of stimulation.]
{{\bf Application of 10 $\mu$M APV, 50 $\mu$M APV, or 10 $\mu$M nimodipine reduced postsynaptic Ca$^{2+}$ induced by different frequencies of stimulation.} {Reduction of [Ca$^{2+}$]$_{i}$ signals was observed in the soma, the proximal dendrite, and the dendrite at mid-radiatum. (n = 3 -- 7 for each point) }}
\label{Fig7}
\end{figure}

\subsection{Effects of nimodipine or APV on induction of plasticity}

In experiments in which 3~Hz stimulation was applied in the control condition, LTD of about 25--30\% was observed (Fig. 8).  No plasticity was observed when stimuli were applied in the presence of nimodipine or in the presence of 10~$\mu$M or 50~$\mu$M APV.  Similarly, 10~Hz stimulation in the control condition gave rise to LTD of about 30\% (Fig. 9).  While 10 and 50~$\mu$M APV blocked the depression induced by 10 Hz stimulation, nimodipine did not.  Application of 30~Hz stimulation gave rise to a potentiation greater than 50\% in the control condition (Fig. 10).  In the presence of nimodipine or 10~$\mu$M APV, however, no potentiation was observed.  Rather, there was a depression of about 50\%.  In the presence of 50~$\mu$M APV, no plasticity was observed.  

\begin{figure}[p]
\centerline {\epsfxsize=6.5in 
\epsffile{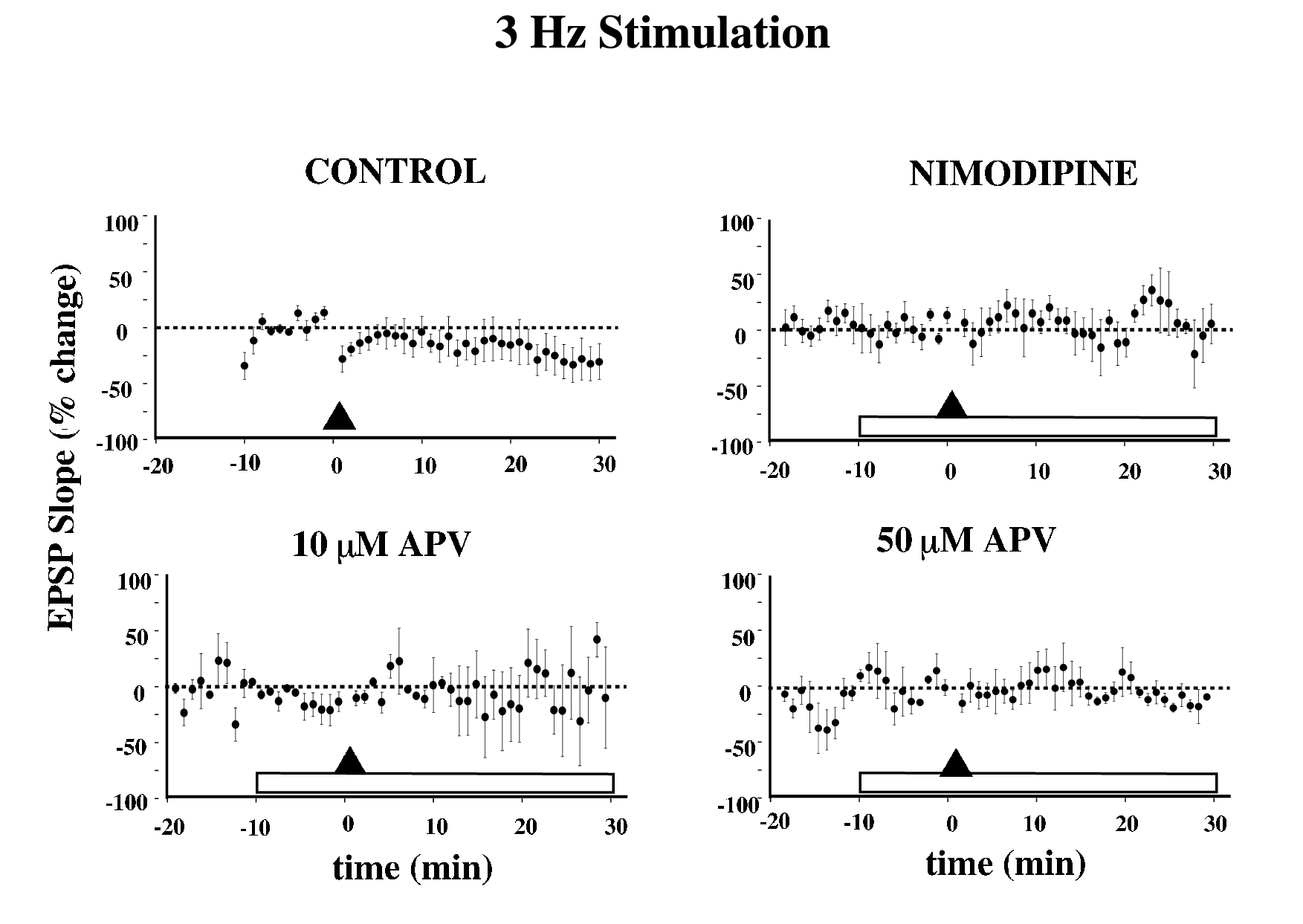}}

\caption[Addition of nimodipine, 10 $\mu$M APV, or 50 $\mu$M APV altered the synaptic plasticity induced with 3 Hz stimulation.]
{{\bf Addition of nimodipine, 10 $\mu$M APV, or 50 $\mu$M APV altered the synaptic plasticity induced with 3 Hz stimulation.} {Plots represent group data for experiments in which 3 Hz stimuli were applied, and data are expressed as percent change in response size in the control condition, in the presence of nimodipine, and in the presence of  10 and 50 $\mu$M APV.  Triangles indicate the stimulation interval, defined as time 0, and bars indicate the period of drug application.  In control experiments baseline responses were obtained for 10 min before stimulation was delivered.  In experiments where drugs were applied, an additional 10 min of baseline responses were obtained in the presence of the drug prior to the delivery of stimuli.  Responses were recorded for at least 30 min following the stimulation.   3 Hz stimulation resulted in LTD in the control condition.  No plasticity was observed in the presence of nimodipine, 10 $\mu$M APV, or 50 $\mu$M APV. (n = 3 -- 7 for each condition)}}
\label{Fig8}
\end{figure}

\begin{figure}[p]
\centerline {\epsfxsize=6.5in 
\epsffile{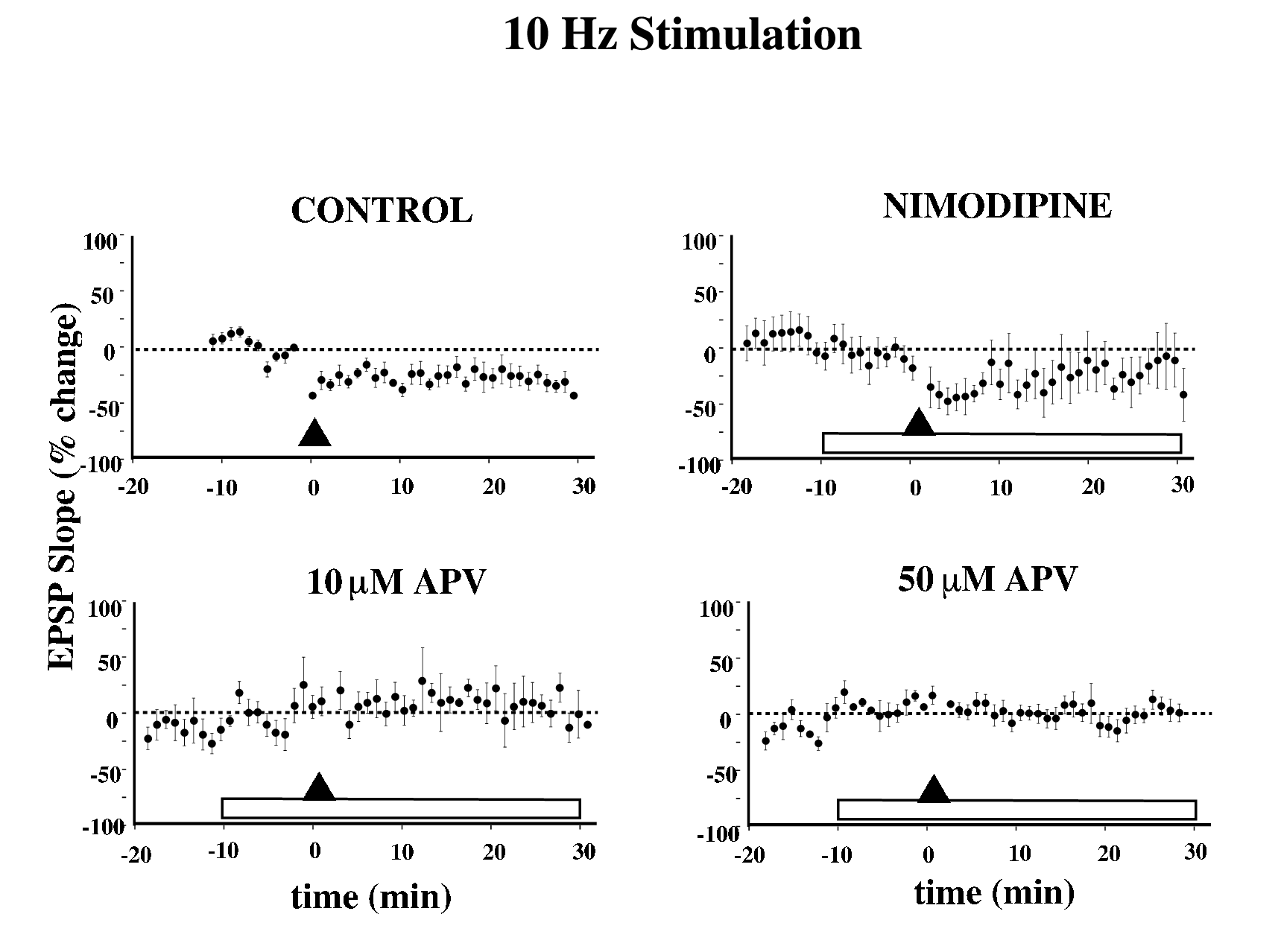}}

\caption[Addition of nimodipine, 10 $\mu$M APV, or 50 $\mu$M APV altered the synaptic plasticity induced with 10 Hz stimulation.]
{{\bf Addition of nimodipine, 10 $\mu$M APV, or 50 $\mu$M APV altered the synaptic plasticity induced with 10 Hz stimulation.} {10 Hz stimulation resulted in LTD in the control condition and in the presence of nimodipine.  No plasticity was observed in the presence of 10 $\mu$M APV or 50 $\mu$M APV. (n = 3 -- 7 for each condition)}}
\label{Fig9}
\end{figure}

\begin{figure}[p]
\centerline {\epsfxsize=6.5in 
\epsffile{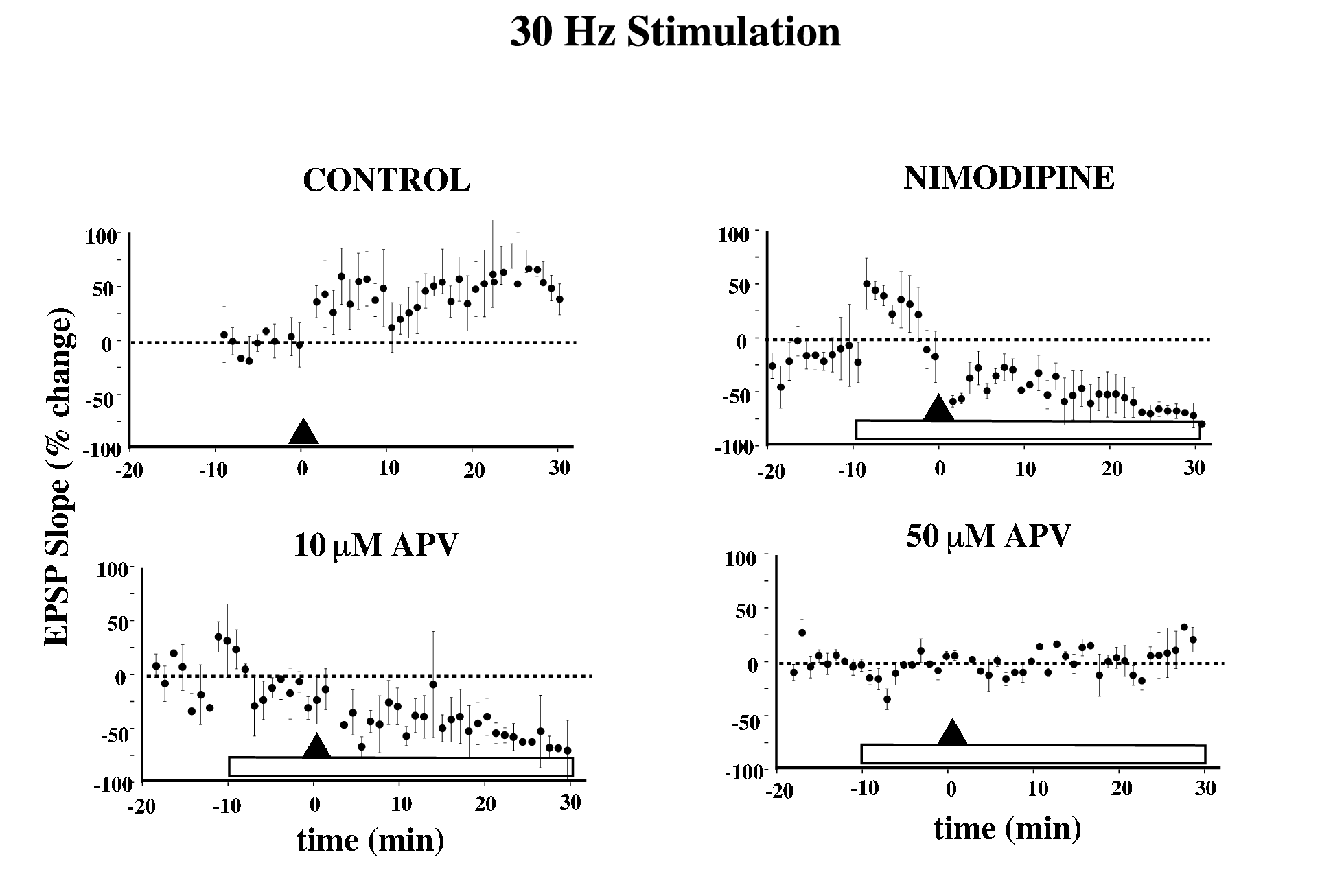}}

\caption[Addition of nimodipine, 10 $\mu$M APV, or 50 $\mu$M APV altered the synaptic plasticity induced with 30 Hz stimulation.]
{{\bf Addition of nimodipine, 10 $\mu$M APV, or 50 $\mu$M APV altered the synaptic plasticity induced with 30 Hz stimulation.} {30 Hz stimulation resulted in LTP in the control condition.  LTD was observed in the presence of nimodipine or 10 $\mu$M APV, and no plasticity was observed in the presence of 50 $\mu$M APV.  (n = 3 -- 7 for each condition)}}
\label{Fig10}
\end{figure}

Like 30~Hz stimulation, 50~Hz stimulation also gave rise to potentiation in the control condition (Fig. 11).  Again, application of nimodipine or 10~$\mu$M APV converted this potentiation into a depression, while in the presence of 50~$\mu$M APV, no plasticity was observed. With 100~Hz stimulation LTP was observed in control conditions (Fig. 12).  In the presence of nimodipine a small depression was observed, while no plasticity was obtained with either 10 or 50~$\mu$M APV.   With 200~Hz stimulation potentiation was again observed in control conditions, there was no plasticity with nimodipine or 50~$\mu$M APV, but potentiation was observed in the presence of 10~$\mu$M APV (Fig 13).

\begin{figure}[p]
\centerline {\epsfxsize=6.5in 
\epsffile{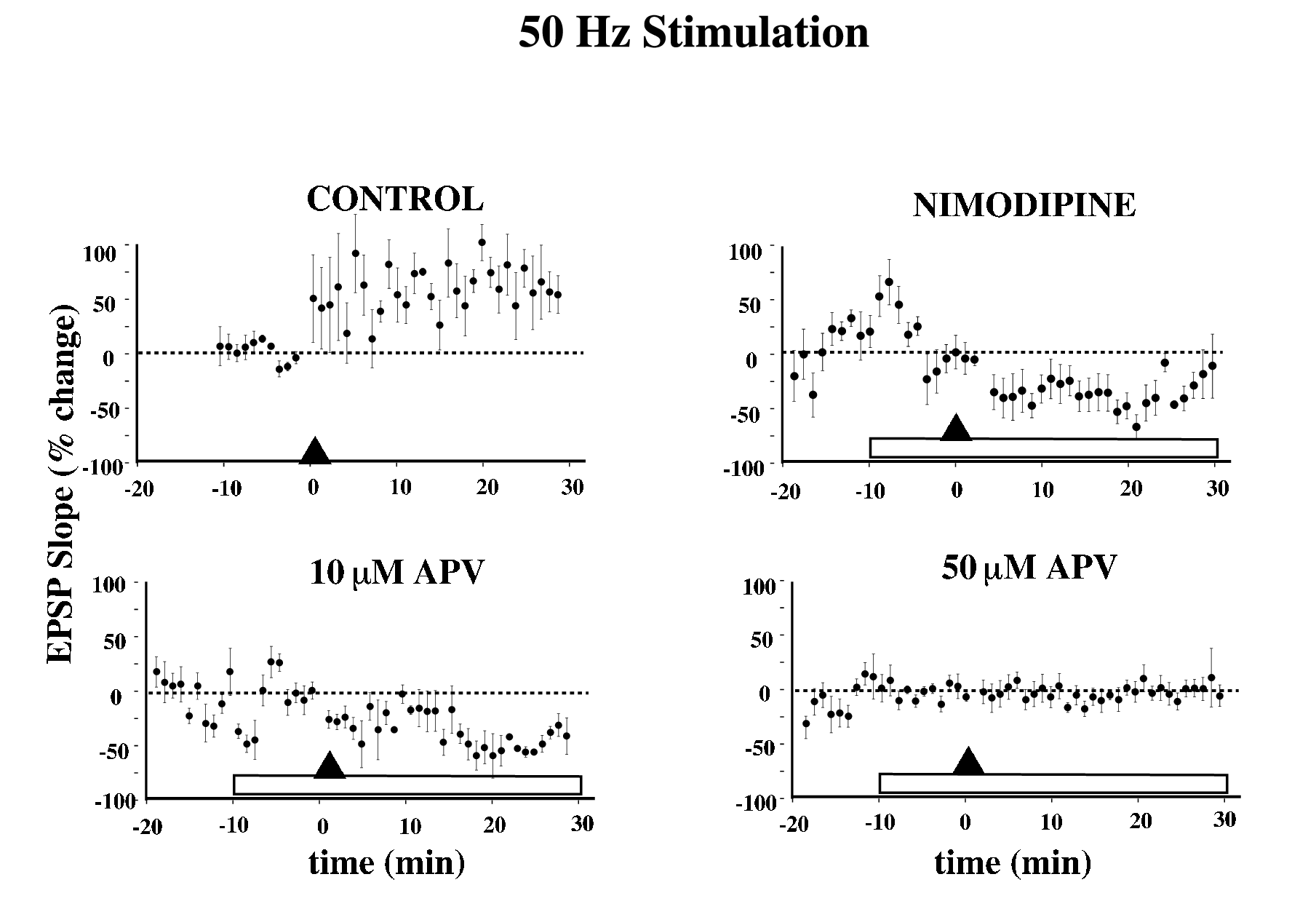}}

\caption[Addition of nimodipine, 10 $\mu$M APV, or 50 $\mu$M APV altered the synaptic plasticity induced with 50 Hz stimulation.]
{{\bf Addition of nimodipine, 10 $\mu$M APV, or 50 $\mu$M APV altered the synaptic plasticity induced with 50 Hz stimulation.} {50 Hz stimulation resulted in LTP in the control condition.  LTD was observed in the presence of nimodipine or 10 $\mu$M APV, and no plasticity was observed in the presence of 50 $\mu$M APV.   (n = 3 -- 7 for each condition)}}
\label{Fig11}
\end{figure}

\begin{figure}[p]
\centerline {\epsfxsize=6.5in 
\epsffile{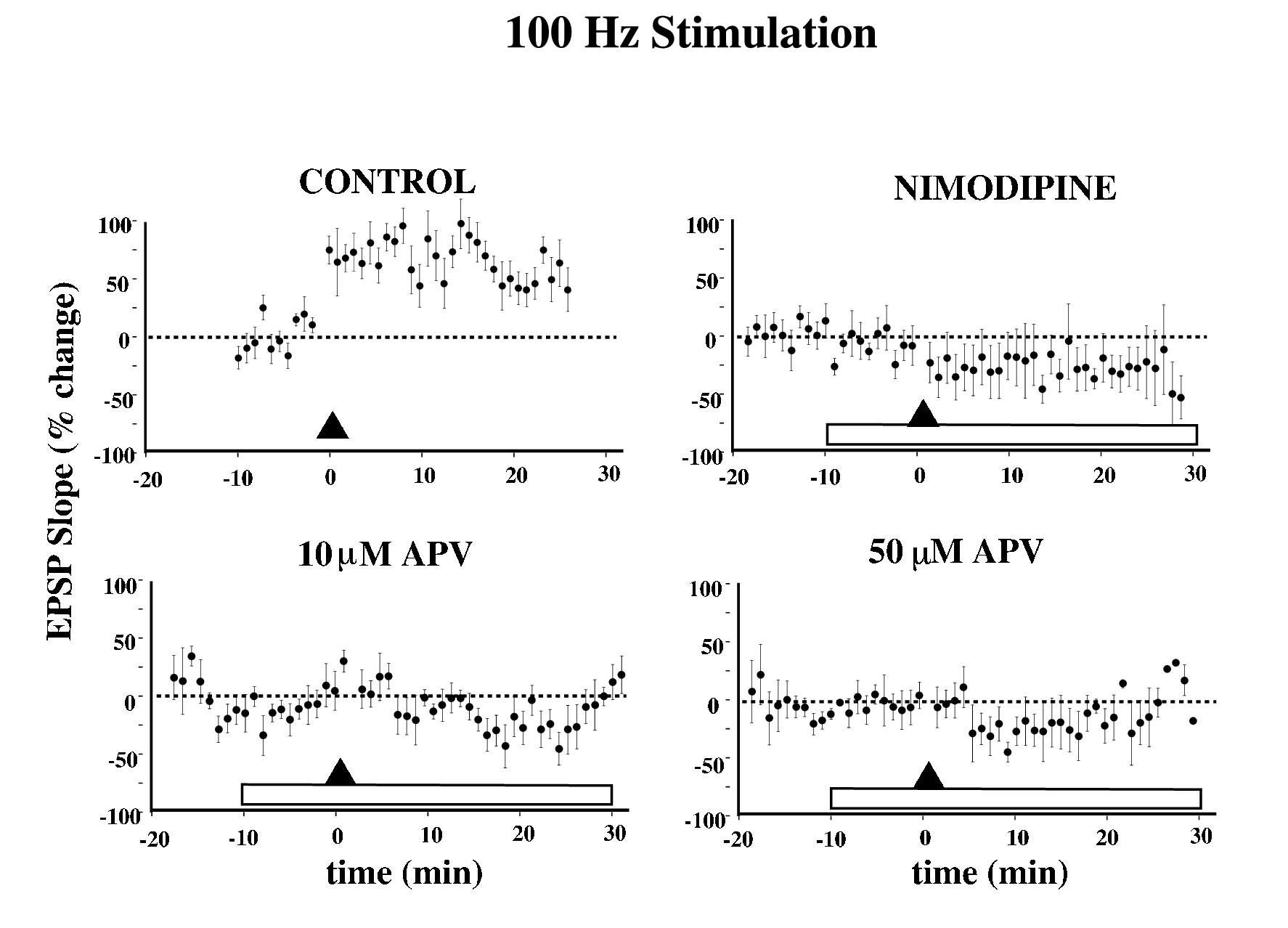}}

\caption[Addition of nimodipine, 10 $\mu$M APV, or 50 $\mu$M APV altered the synaptic plasticity induced with 100 Hz stimulation.]
{{\bf Addition of nimodipine, 10 $\mu$M APV, or 50 $\mu$M APV altered the synaptic plasticity induced with 100 Hz stimulation.} {100 Hz stimulation resulted in LTP in the control condition, LTD was observed in the presence of nimodipine, and only transient depression was observed in the presence of 10 $\mu$M APV or 50 $\mu$M APV. (n = 3 -- 7 for each condition)}}
\label{Fig12}
\end{figure}

\begin{figure}[p]
\centerline {\epsfxsize=6.5in 
\epsffile{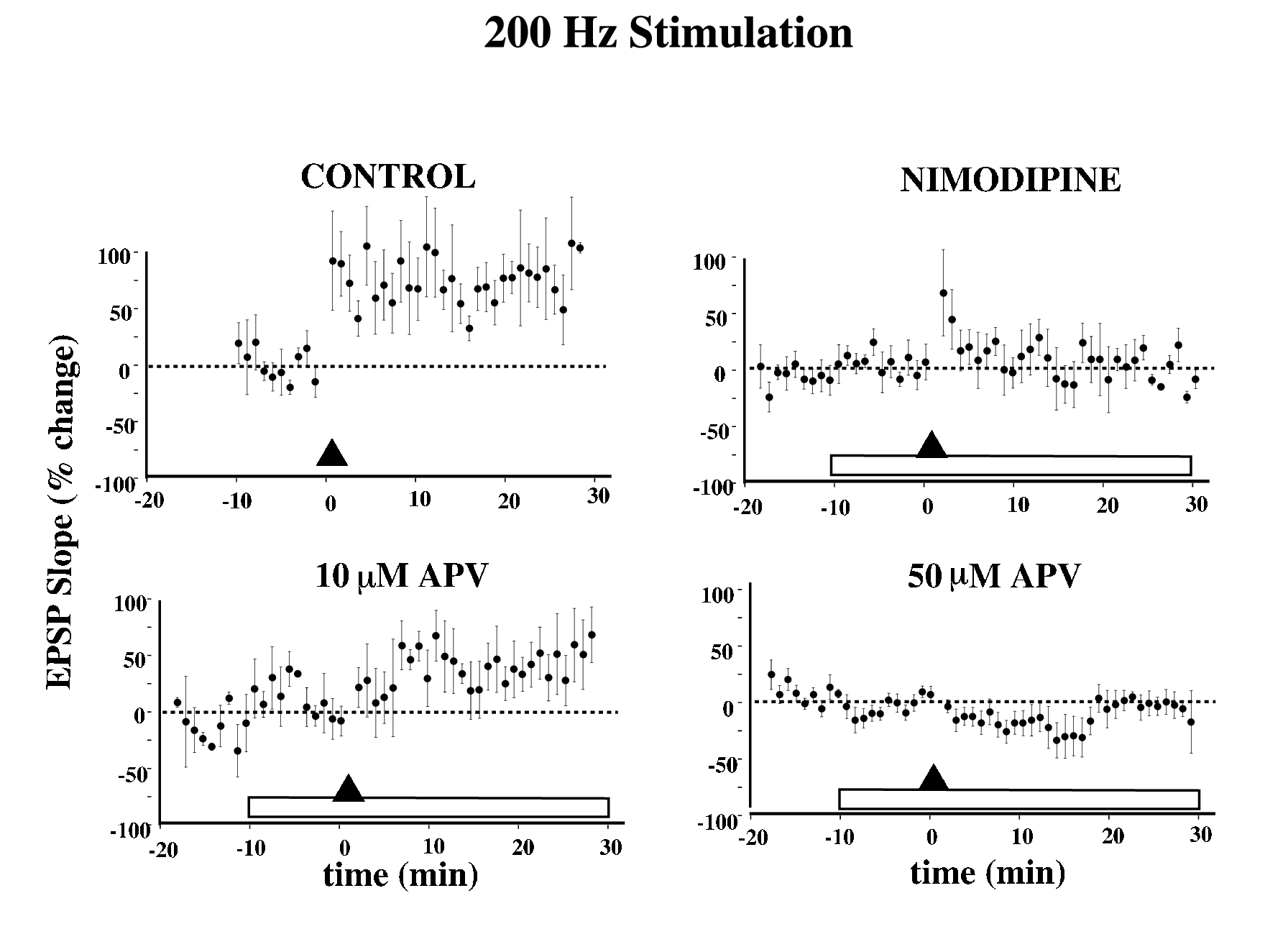}}

\caption[Addition of nimodipine, 10 $\mu$M APV, or 50 $\mu$M APV altered the synaptic plasticity induced with 200 Hz stimulation.]
{{\bf Addition of nimodipine, 10 $\mu$M APV, or 50 $\mu$M APV altered the synaptic plasticity induced with 200 Hz stimulation.} {200 Hz stimulation resulted in LTP in the control condition.  No plasticity was observed in the presence of nimodipine, reduced potentiation was observed in the presence of 10 $\mu$M APV, and only transient depression was observed in the presence of 50 $\mu$M APV.  (n = 3 -- 7 for each condition)}}
\label{Fig13}
\end{figure}

\subsection{Changes in the LTD/LTP transition frequency with reductions in postsynaptic [Ca$^{2+}$]$_{i}$}

The changes in synaptic strength illustrated in Figures 8-13 were plotted as a function of stimulus frequency for control and in the presence of each blocker (Fig. 14) .  In control conditions depression was observed at 3 and 10 Hz while potentiation was observed at 30, 50, 100 and 200~Hz.  In the presence of nimodipine and in the presence of APV, however, the curves were shifted to the right, with the transition from LTD to LTP occurring at a higher frequency.  In the presence of these drugs, 3~Hz stimulation resulted in little or no change in synaptic strength.  With 10~Hz stimulation 10~$\mu$M APV blocked the depression, while nimodipine did not.  In the presence of both 10~$\mu$M APV and nimodipine, 30, 50 and 100~Hz stimulation resulted in depression rather than potentiation.  At 200~Hz, some potentiation was observed in  10~$\mu$M APV, while no plasticity was observed in the presence of nimodipine and only transient depression was observed in 50~$\mu$M APV.  

	The changes in synaptic strength shown in the previous figure are also plotted as a function of postsynaptic [Ca$^{2+}$] for all experiments, binned in intervals of 5\% $\Delta$F/F (Fig. 15).  The relationship between plasticity and [Ca$^{2+}$]$_{i}$ is shown for each of the 3 regions of the neuron.  In each region of the neuron there was no appreciable change in synaptic strength where the postsynaptic Ca$^{2+}$ signal induced by the stimulation was lowest.  Depression was observed where signals were at an intermediate value, and where Ca$^{2+}$ levels were highest there was potentiation. In the soma the APV experiments are plotted separately (dashed line) to illustrate that there was no relationship between [Ca$^{2+}$]$_{i}$ and plasticity in this region.

\begin{figure}[p]
\centerline {\epsfxsize=6.5in 
\epsffile{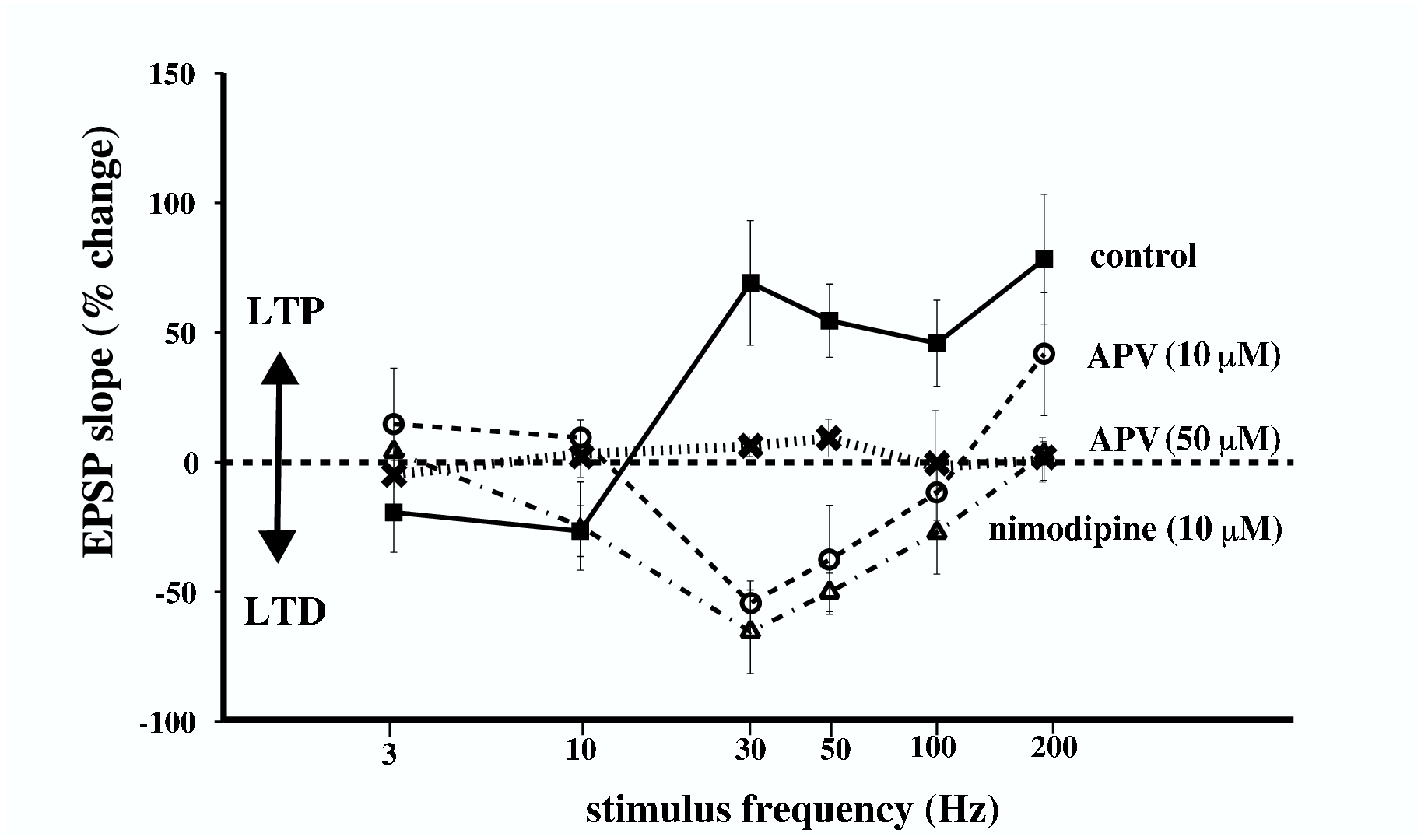}}

\caption[Changes in synaptic strength as a function of stimulus frequency.]
{{\bf Changes in synaptic strength as a function of stimulus frequency.} {Synaptic plasticity values are shown for the control condition, and in the presence of nimodipine, 10 $\mu$M APV, or 50 $\mu$M APV.  In control conditions depression was observed at 3 and 10 Hz while potentiation was observed at 30, 50, 100, and 200~Hz.  Curves were shifted to the right in the presence of nimodipine and in the presence of 10 $\mu$M APV such that the transition from LTD to LTP occurred at a higher frequency.  In the presence of 50 $\mu$M APV no plasticity was observed at any frequency. (n = 3 -- 7 for each point)}}
\label{Fig14}
\end{figure}

\begin{figure}[p]
\centerline {\epsfxsize=3in 
\epsffile{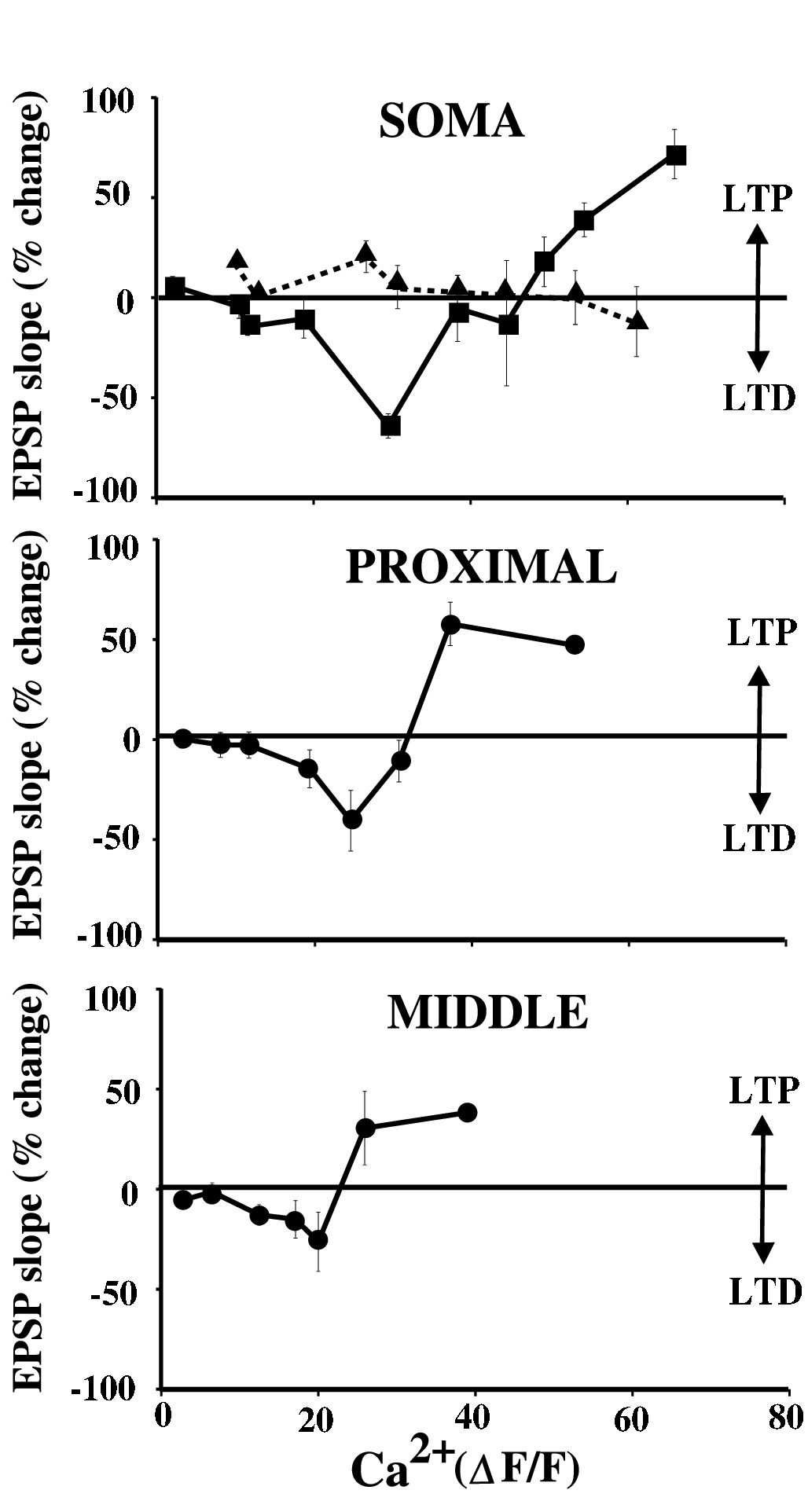}}

\caption[Relationship between synaptic plasticity and postsynaptic Ca$^{2+}$ for each of the three regions of the neuron.]
{{\bf Relationship between synaptic plasticity and postsynaptic Ca$^{2+}$ for each of the three regions of the neuron.} {Changes in synaptic strength over all experiments are plotted as a function of postsynaptic [Ca$^{2+}$] for the soma and the proximal and the middle dendrites, binned in intervals of 5\% $\Delta$F/F (n = 2 -- 10 for each binned point).  In all three regions of the neuron there is no appreciable change in synaptic strength where the postsynaptic Ca$^{2+}$ signal induced by the stimulation was lowest.  Depression was observed where Ca$^{2+}$ signals were at an intermediate value, and potentiation was observed where Ca$^{2+}$ signals were highest. For the soma, APV data are plotted separately (dashed line), because 10 or 50 $\mu$M APV did not reduce postsynaptic Ca$^{2+}$ signals, though they reduced or blocked plasticity.  When these APV data are plotted separately, the remainder of the data for the soma show a similar relationship as in the dendrites, where lower [Ca$^{2+}$]$_{i}$ led to depression and higher [Ca$^{2+}$]$_{i}$ led to potentiation.}}
\label{Fig15}
\end{figure}

\section{Ca$^{2+}$ Imaging Reveals Localized Increases in Dendritic Action Potential Amplitude Following LTP Induction}

We tested the hypothesis that LTP induction results in localized decreases in dendritic K$^{+}$ channel activity and increases in amplitude of back-propagating action potentials. We attempted to observe these increases in amplitude indirectly as localized increases in dendritic Ca$^{2+}$signals following LTP induction. This method eliminated the need to record at the precise site of the synapse to detect the increases in action potential amplitude.  We filled cells with bis-fura for fluorescence imaging of postsynaptic Ca$^{2+}$ in the dendrites to determine whether there were localized increases in postsynaptic Ca$^{2+}$ following high-frequency pairing stimulation at 100 Hz (HFS). It was possible to image intracellular Ca$^{2+}$ throughout the dendritic tree during the application of a train of 5 back-propagating action potentials applied at 20 Hz (Fig. 16). The changes in postsynaptic Ca$^{2+}$ induced by the stimulation were measured as percent $\Delta$F/F for 25 $\mu$m segments of the dendritic tree at distances of 100 to 300 $\mu$m, averaged over 15 traces. A train of action potentials was applied before the application of HFS, 5 min after the application of HFS, and 30 min after the application of HFS for the purposes of measuring postsynaptic Ca$^{2+}$ signals. Percent change in $\Delta$F/F was calculated for each region of the dendrite at each time point. 

	It has been shown that subthreshold stimuli preceded by a hyperpolarizing prepulse give rise to very localized increases in Ca$^{2+}$at the synapse [. magee subthreshold @13563@.]. This technique was used in conjunction with the above experiments to determine the site(s) of synaptic input and whether they correspond to the site(s) of increased Ca$^{2+}$ influx. At the beginning of each experiment, a train of 5 subthreshold EPSPs of 20-30 mV at 50 Hz was applied 400 ms after a 1500 ms hyperpolarizing prepulse of 15--25 mV, and changes in dendritic Ca$^{2+}$ were measured as percent $\Delta$F/F for 25 $\mu$m segments of the dendrite at distances of 100 to 300 $\mu$m, averaged over 15 traces (Fig. 17). The region of the dendrite that showed the largest Ca$^{2+}$ signal in response to the subthreshold EPSPs was taken to be the region of synaptic input.

\begin{figure}[p]
\centerline {\epsfxsize=6in 
\epsffile{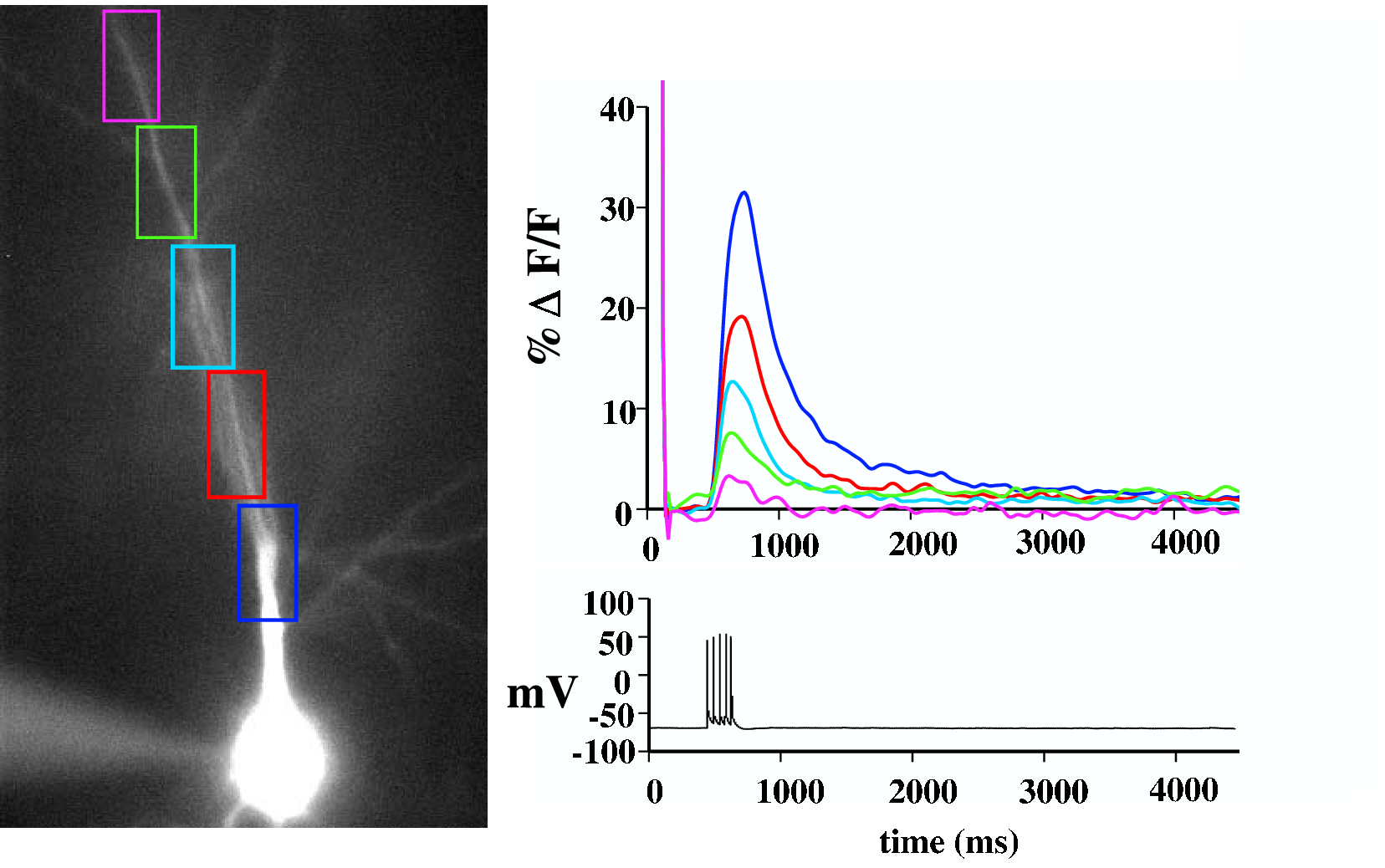}}

\caption[Sample fluorescence traces during a train of back-propagating action potentials.]
{{\bf Sample fluorescence traces during a train of back-propagating action potentials.} {The changes in postysynaptic Ca$^{2+}$ induced by a brief train of action potentials were measured as percent $\Delta$F/F for 25 $\mu$m segments of the dendritic tree at distances of 100 to 300 $\mu$m, averaged over 15 traces. A train of 5 back-propagating action potentials applied at 20 Hz was applied before the application of HFS, 5 min after the application of HFS, and 30 min after the application of HFS for the purposes of measuring postsynaptic Ca$^{2+}$ signals. Percent change in $\Delta$F/F was calculated for each region of the dendrite at each time point.}}
\label{Fig16}
\end{figure}

\begin{figure}[p]
\centerline {\epsfxsize=6in 
\epsffile{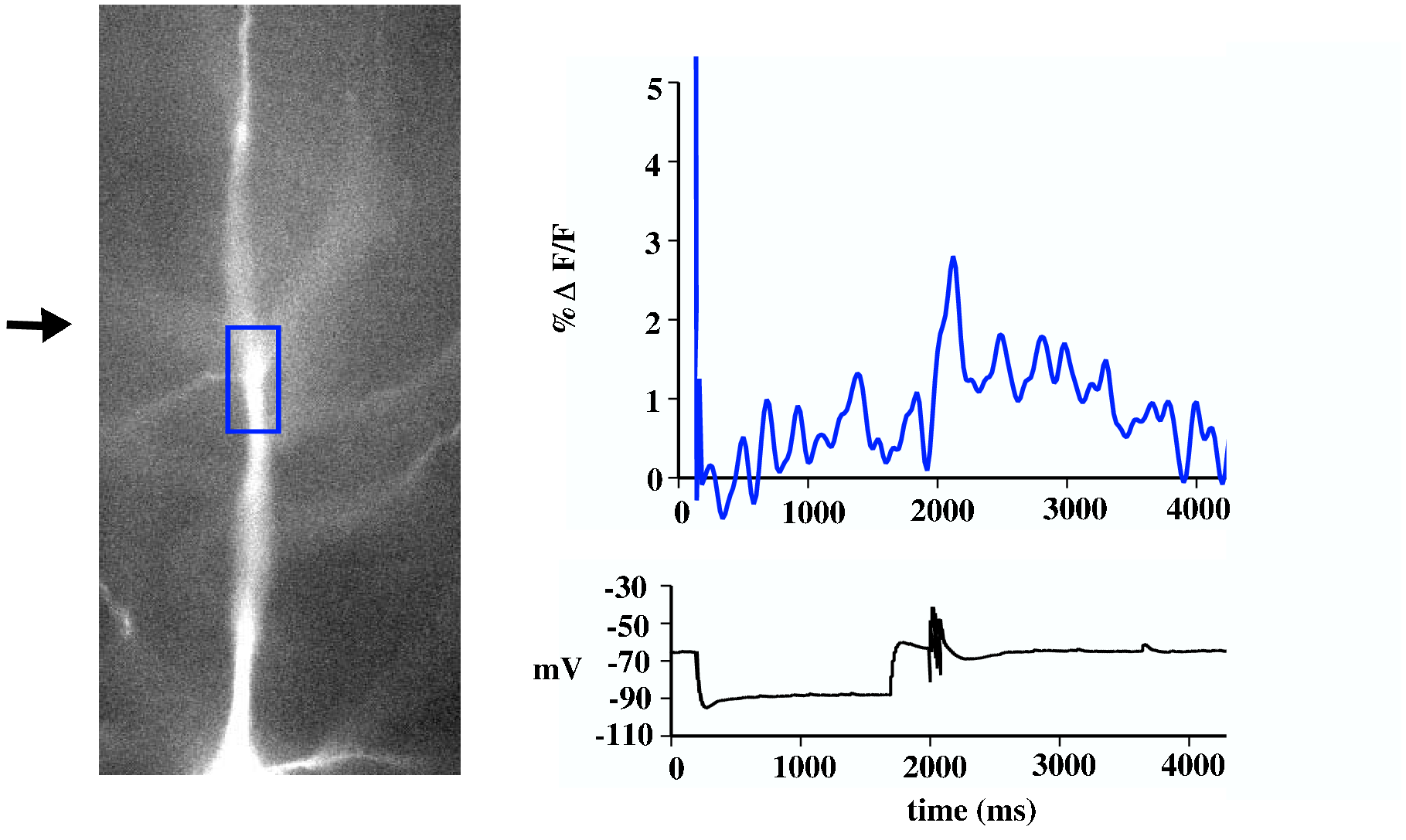}}

\caption[Sample fluorescence traces during a train of subthreshold EPSPs.]
{{\bf Sample fluorescence traces during a train of subthreshold EPSPs.} {At the beginning of each experiment, a train of 5 subthreshold EPSPs of 20--30 mV at 50 Hz was applied 400 ms after a 1500 ms hyperpolarizing prepulse of 15--25 mV, and changes in dendritic Ca$^{2+}$ were measured as percent $\Delta$F/F for 25 $\mu$m segments of the dendrite at distances of 100 to 300 $\mu$m, averaged over 15 traces. The region of the dendrite that showed the largest Ca$^{2+}$ signal in response to the subthreshold EPSPs (boxed region) was taken to be the region of synaptic input.  The position of the stimulating electrode is indicated by an arrow.}}
\label{Fig17}
\end{figure}

\subsection{Increases in dendritic Ca$^{2+}$ signals following LTP}

When 100 Hz pairing stimulation was applied, LTP was consistently induced and was measured at 30 min post-HFS (198.75 $\pm$ 46.4\%, n=8, p<0.005).  In these experiments there were significant increases in dendritic Ca$^{2+}$ signals at the region of the synapse 5 min after the application of HFS (20.3 $\pm$ 4.3\%, n=8, p<0.01)  Fig. 18). Significant but smaller increases in the adjacent 25 $\mu$m segment of the dendrite were also observed (8.1 $\pm$ 2.8\%, n=8, p<0.05). Other regions of the dendrite showed little increase in the size of Ca$^{2+}$ signals after application of HFS. In addition, at 30 min after HFS application there were no consistent increases in the size of dendritic Ca$^{2+}$ signals at the synapse or in any region.

\begin{figure}[p]
\centerline {\epsfxsize=6in 
\epsffile{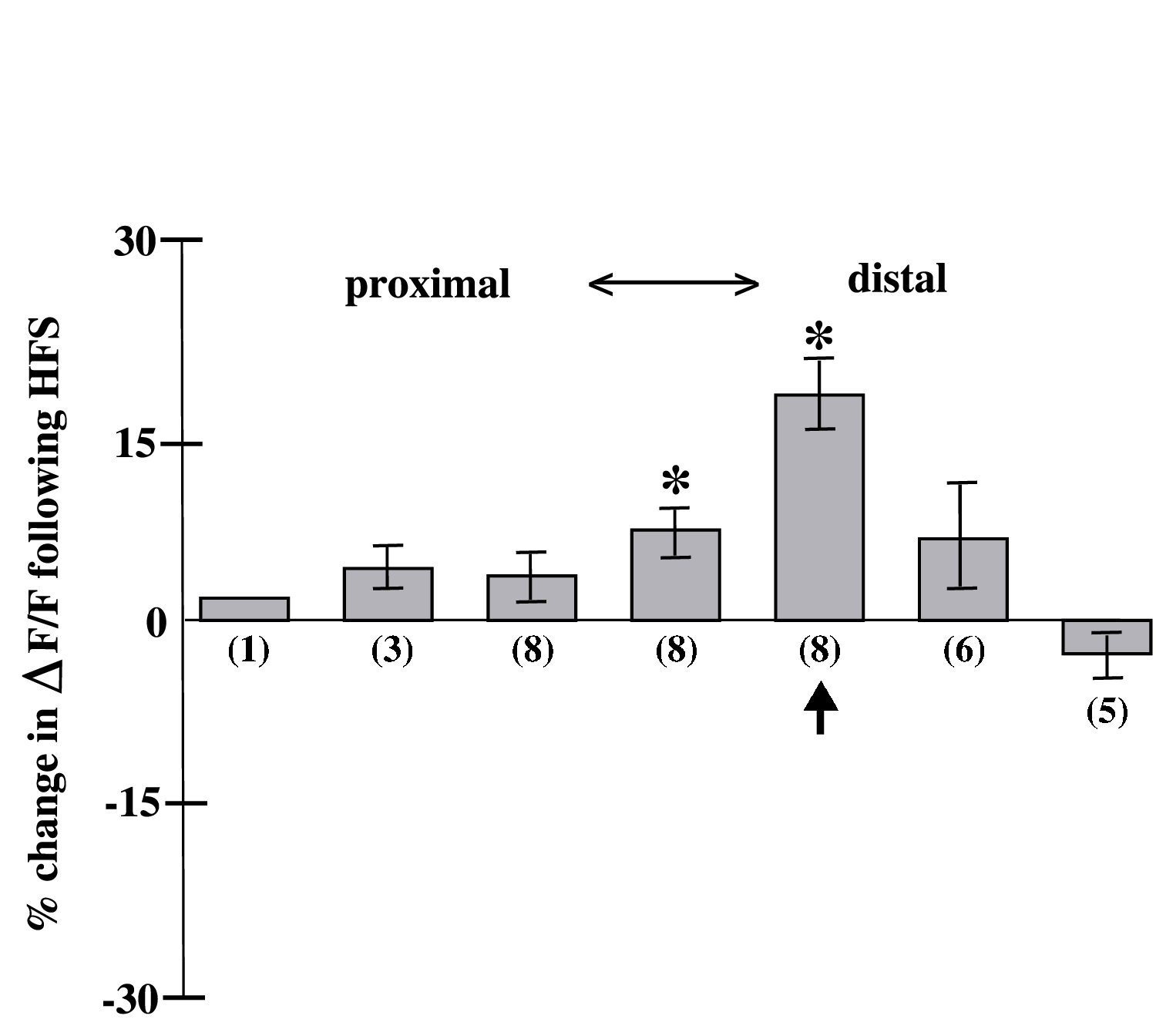}}

\caption[Changes in dendritic Ca$^{2+}$ signals following 100 Hz pairing stimulation.]
{{\bf Changes in dendritic Ca$^{2+}$ signals following 100 Hz pairing stimulation.} {Histograms represent percent change in $\Delta$F/F $\pm$SEM for adjacent 25 $\mu$m segments of the dendrite.  The segment of the dendrite corresponding to the location of the synapse is marked with an arrow.  Distances were normalized with respect to the location of the synapse and percent change in $\Delta$F/F was averaged over all experiments for each region.  In experiments in which LTP was induced by 100 Hz pairing stimulation, there were consistent increases in dendritic Ca$^{2+}$ signals at the region of the synapse  5 min after the application of HFS (20.3 $\pm$4.3\%, n=8, p<0.01).  Significant but smaller increases in the adjacent 25 $\mu$m segment of the dendrite were also observed (8.1 $\pm$ 2.8\%, n=8, p<0.05).  Significant changes are noted with an asterisk.  Number of observations (n) for each region is noted in parentheses.}}
\label{Fig18}
\end{figure}

\subsection{Increases in dendritic Ca$^{2+}$ signals depend on synaptic input and NMDA receptors}

It is known that back-propagating action potentials do not produce LTP in the absence of synaptic stimulation [.magee Hebbian @14316@.]. The above experiments were repeated with 100 Hz back-propagating action potentials only, and no LTP was induced. As above, dendritic Ca$^{2+}$ was measured in response to a train of action potentials applied before the application of HFS, 5 min after the application of HFS, and 30 min after the application of HFS. In these experiments, no increases in dendritic Ca$^{2+}$ signals were observed at the synapse following application of HFS, but significant decreases were seen at the synapse and adjacent 25 $\mu$m segment (-16.4 $\pm$ 5.0, n=5, p<0.05, -14.8 $\pm$ 3.2, n=5, p<0.01, Fig. 19). Experiments were also performed in the presence of 50~$\mu$M APV, which consistently blocked LTP induction following 100 Hz pairing stimulation. Again, no increase in dendritic Ca$^{2+}$ signals was observed in the region of the synapse (Figure 20). 

\begin{figure}[p]
\centerline {\epsfxsize=6in 
\epsffile{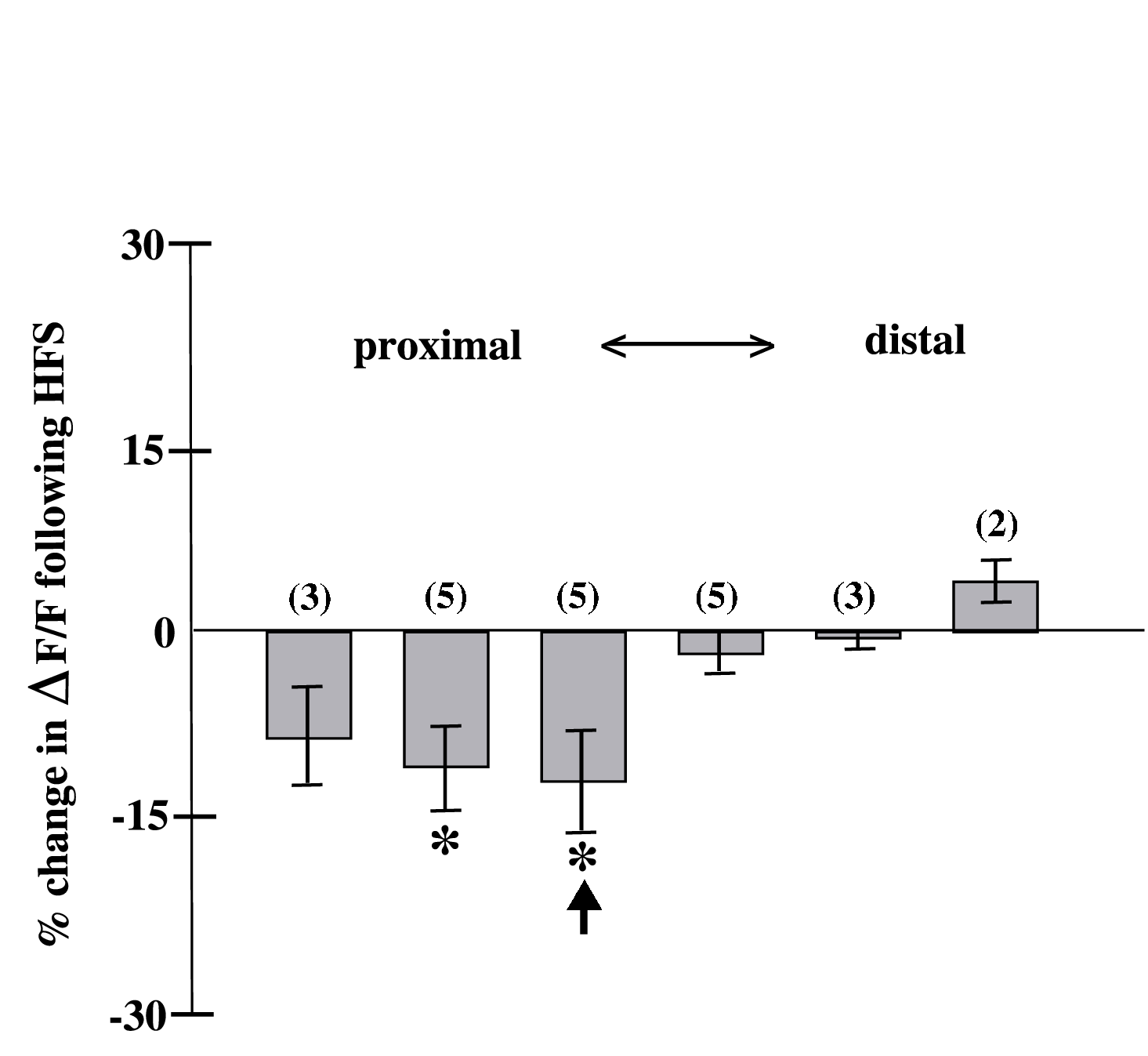}}

\caption[Changes in dendritic Ca$^{2+}$ signals following 100 Hz back-propagating action potentials in the absence of synaptic stimulation.]
{{\bf Changes in dendritic Ca$^{2+}$ signals following 100 Hz back-propagating action potentials in the absence of synaptic stimulation.} {In these experiments, no increases in dendritic Ca$^{2+}$ signals were observed at the synapse following application of HFS but significant decreases were seen at the synapse and adjacent 25 $\mu$m segment (-16.4 $\pm$ 5.0, n=5, p<0.05 and -14.8 $\pm$ 3.2, n=5, p<0.01).  Significant changes are noted with an asterisk.  Number of observations (n) for each region is noted in parentheses.}}
\label{Fig19}
\end{figure}

\begin{figure}[p]
\centerline {\epsfxsize=6in 
\epsffile{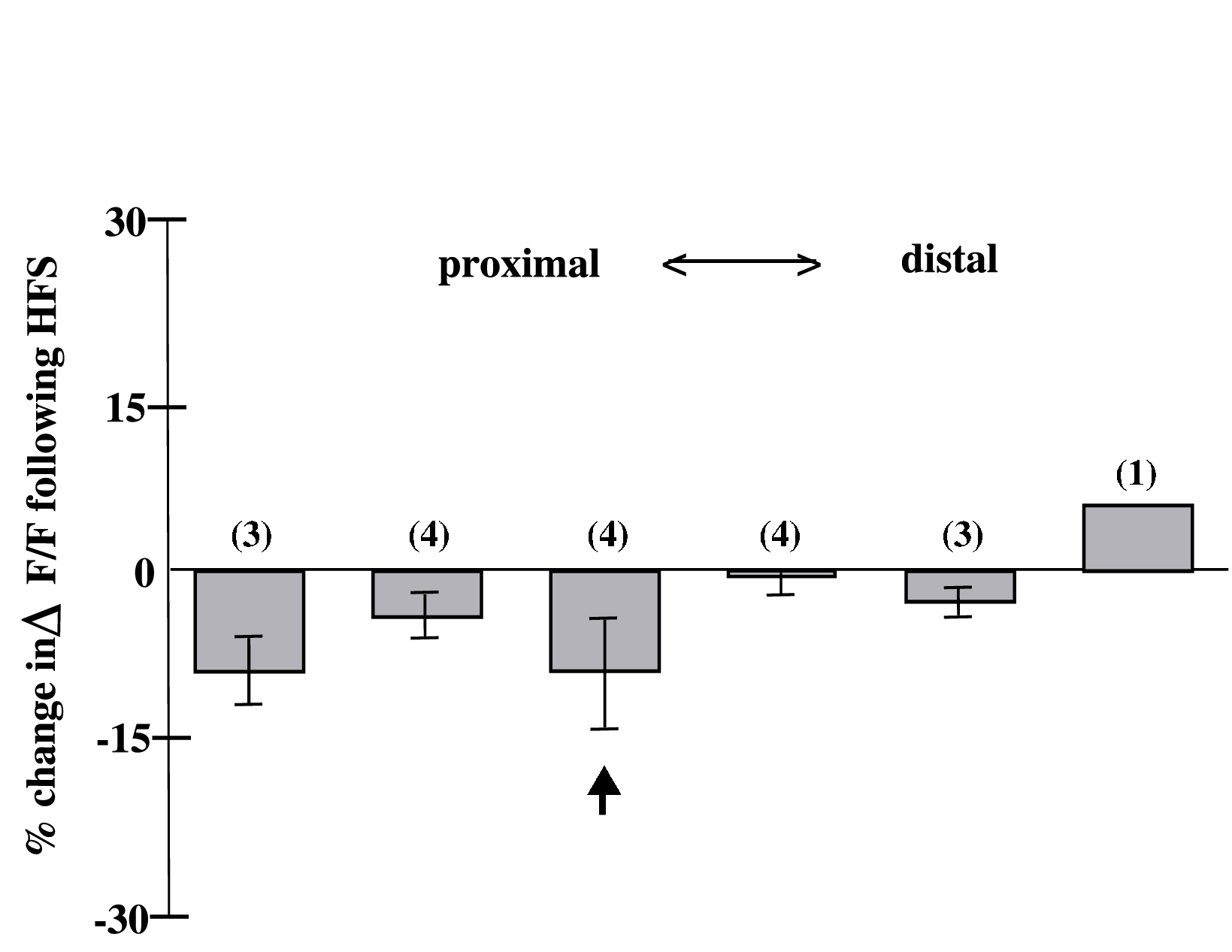}}

\caption[  Changes in dendritic Ca$^{2+}$ signals following 100 Hz pairing stimulation in the presence of 50 $\mu$M APV.]
{{\bf   Changes in dendritic Ca$^{2+}$ signals following 100 Hz pairing stimulation in the presence of 50 $\mu$M APV.} {No increase in dendritic Ca$^{2+}$ signals was observed in the region of the synapse when HFS was administered in the presence of 50 $\mu$M APV.  Number of observations (n) for each region is noted in parentheses.}}
\label{Fig20}
\end{figure}

\subsection{Increases in dendritic Ca$^{2+}$ signals depend on protein kinase activity}

The kinase inhibitor H7 is known to be an effective broad-spectrum inhibitor of protein kinase activity. When the above experiments were performed with 300~$\mu$M H7 included in the the bath, no LTP was induced. In addition, no increases in dendritic Ca$^{2+}$ signals were observed in the region of the synapse (Fig. 21). The kinase inhibitor H89 is a potent and specific inhibitor of protein kinase A (PKA) activity. When 10~$\mu$M H89 was included in the bath, LTP of the usual amplitude was induced. However, no increases in dendritic Ca$^{2+}$ signals were observed in the region of the synapse (Fig.22).

\begin{figure}[p]
\centerline {\epsfxsize=6in 
\epsffile{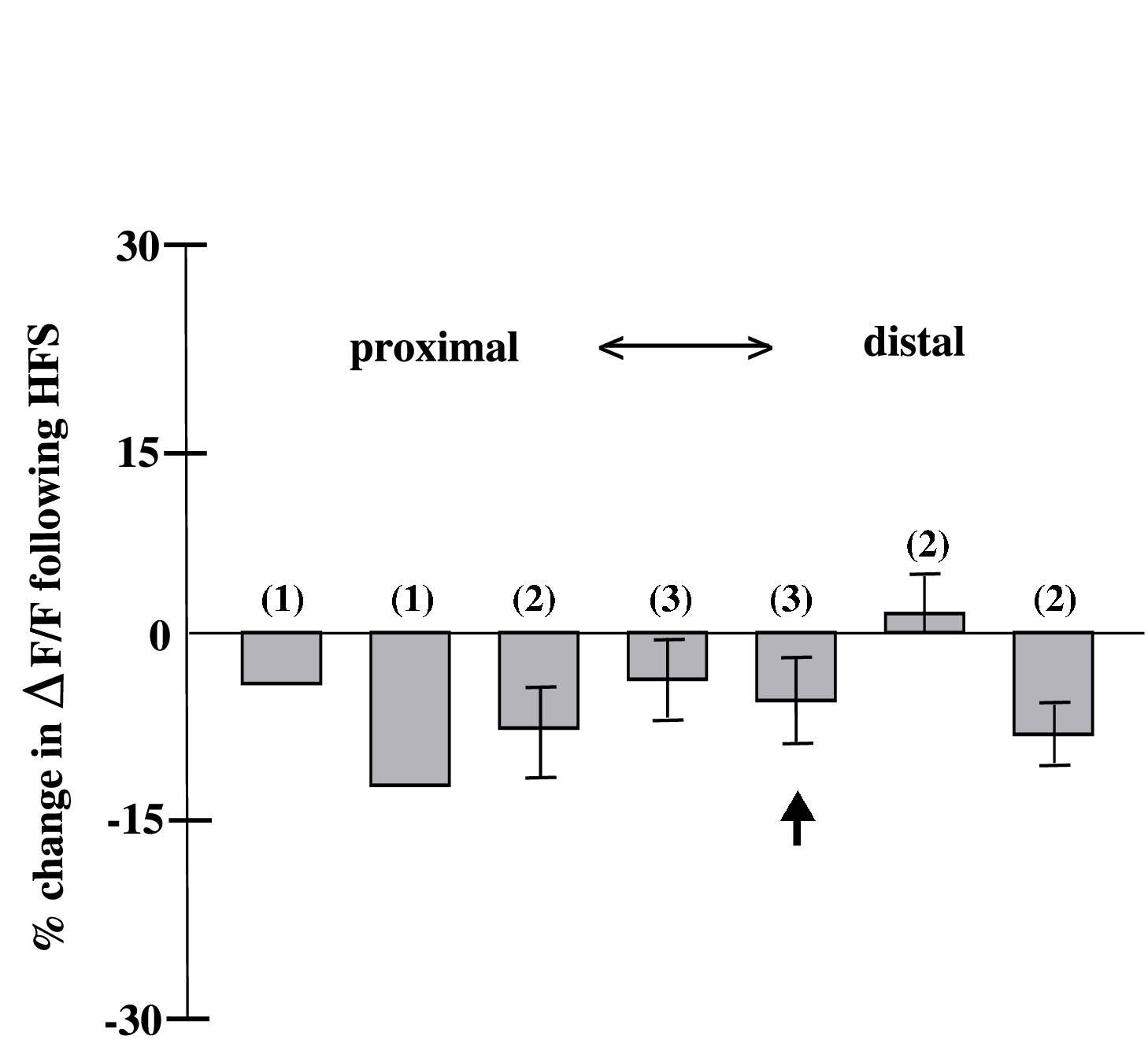}}

\caption[  Changes in dendritic Ca$^{2+}$ signals following 100 Hz pairing stimulation in the presence of 300 $\mu$M H7.]
{{\bf   Changes in dendritic Ca$^{2+}$ signals following 100 Hz pairing stimulation in the presence of 300 $\mu$M H7.} {No increases in dendritic Ca$^{2+}$ signals were observed in the region of the synapse when HFS was administered in the presence of 300 $\mu$M H7.  Number of observations (n) for each region is noted in parentheses. }}
\label{Fig21}
\end{figure}

\begin{figure}[p]
\centerline {\epsfxsize=6in 
\epsffile{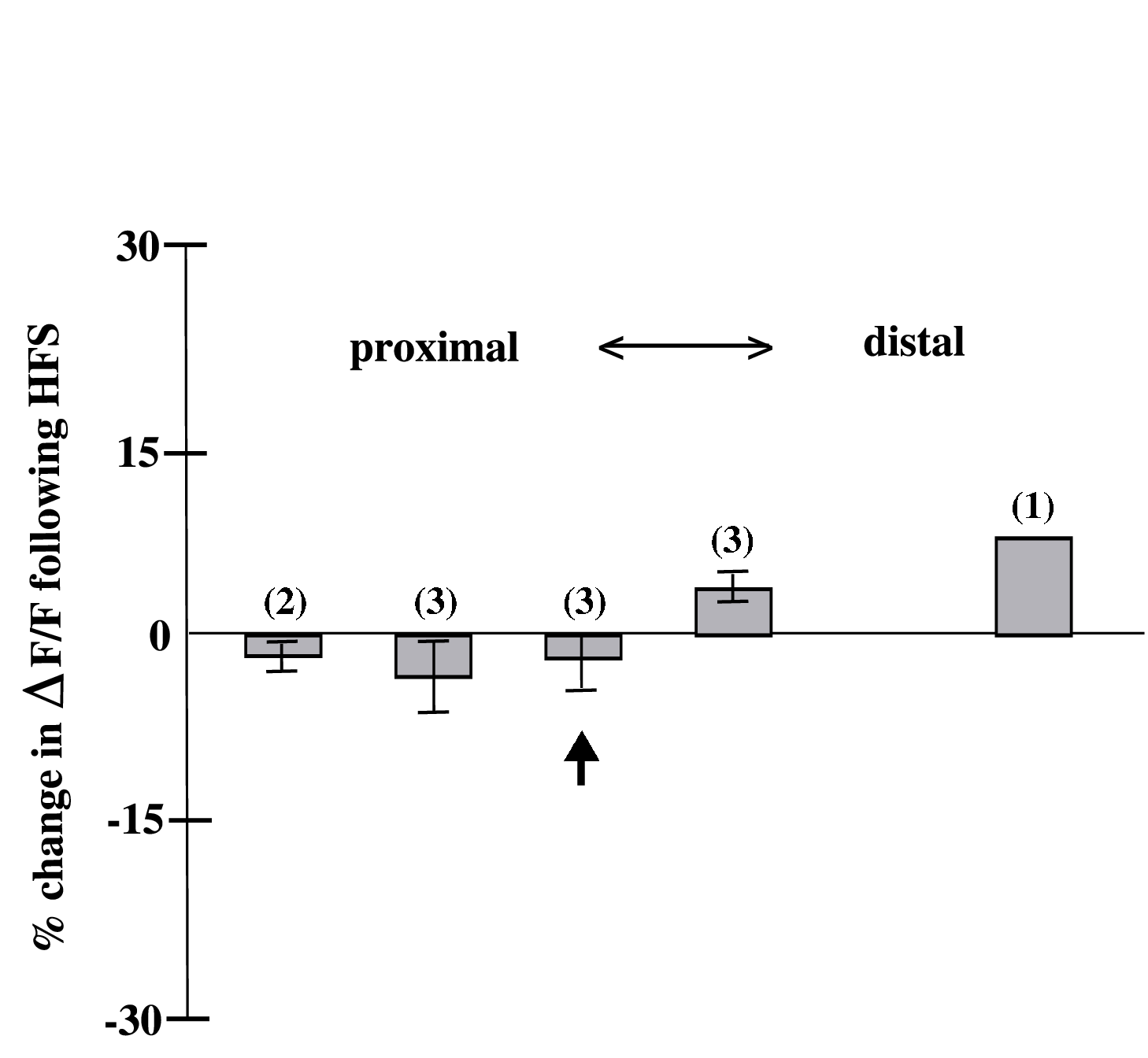}}

\caption[  Changes in dendritic Ca$^{2+}$ signals following 100 Hz pairing stimulation in the presence of 10 $\mu$M H89.]
{{\bf   Changes in dendritic Ca$^{2+}$ signals following 100 Hz pairing stimulation in the presence of 10 $\mu$M H89.} {No increases in dendritic Ca$^{2+}$ signals were observed in the region of the synapse when HFS was administered in the presence of 10 $\mu$M H89.  Number of observations (n) for each region is noted in parentheses.}}
\label{Fig22}
\end{figure}

\begin{figure}[p]
\centerline {\epsfxsize=6in 
\epsffile{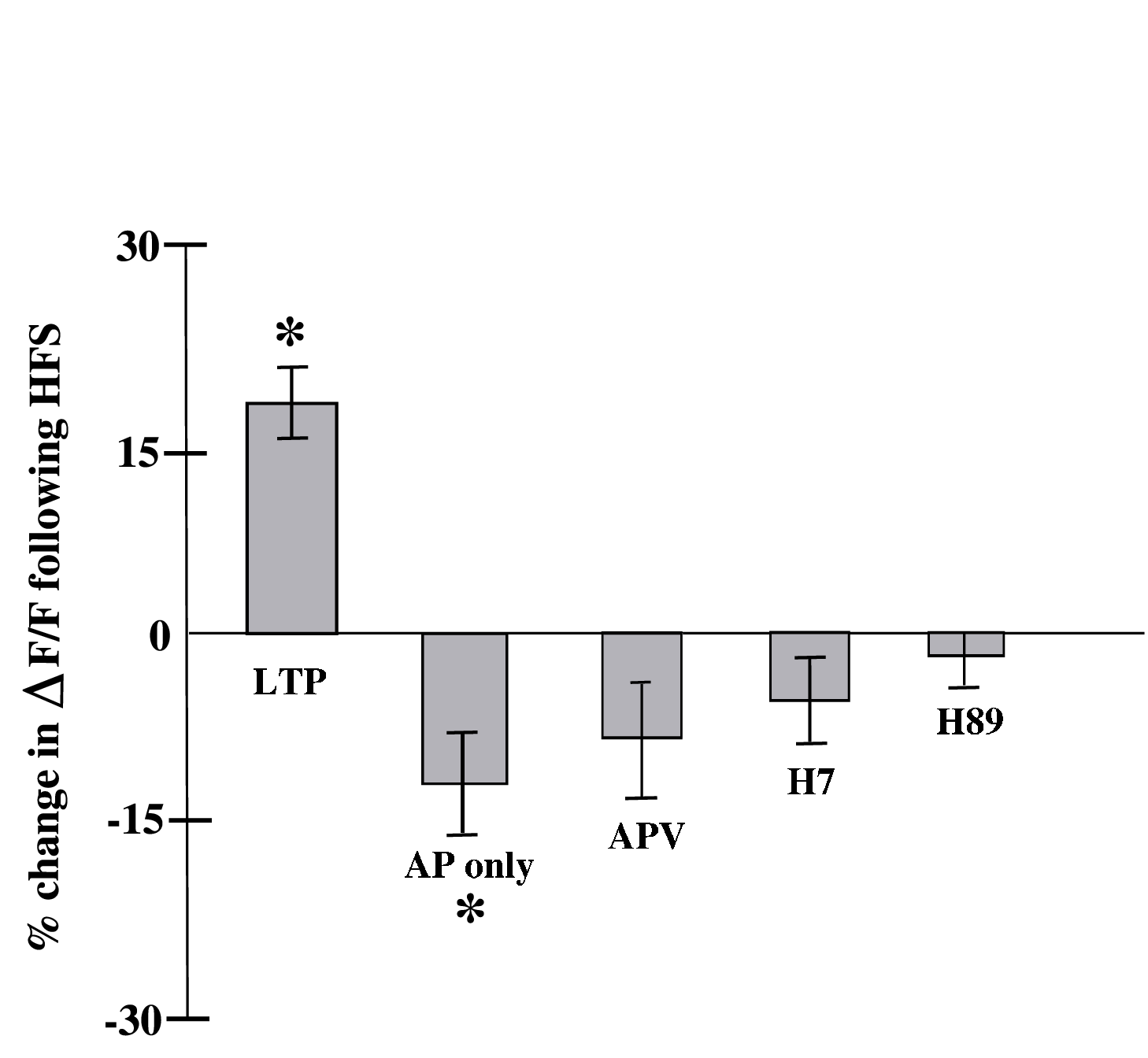}}

\caption[Summary of changes in dendritic Ca$^{2+}$ signals at the synapse for each condition.]
{{\bf  Summary of changes in dendritic Ca$^{2+}$ signals at the synapse for each condition.} {An increase in the size of dendritic Ca$^{2+}$ signals at the synapse following HFS was observed in experiments where LTP was induced under normal conditions.  No increases were seen at the synapse when action potentials were applied without synaptic stimulation, in the presence of APV, H7, or H89. }}
\label{Fig23}
\end{figure}

\subsection{Increases in dendritic Ca$^{2+}$ signals do not correlate with the magnitude of LTP}

To examine the possibility that the size of the increase in dendritic Ca$^{2+}$ signals following HFS corresponds to the magnitude of the LTP observed, the peak increases in dendritic Ca$^{2+}$ observed at the synapse were plotted against the percent LTP for each experiment observed 30 min post-HFS (Fig. 24). There appears to be no correlation between the size of the increase in dendritic Ca$^{2+}$ signal and the size of the LTP. 

\begin{figure}[p]
\centerline {\epsfxsize=6in 
\epsffile{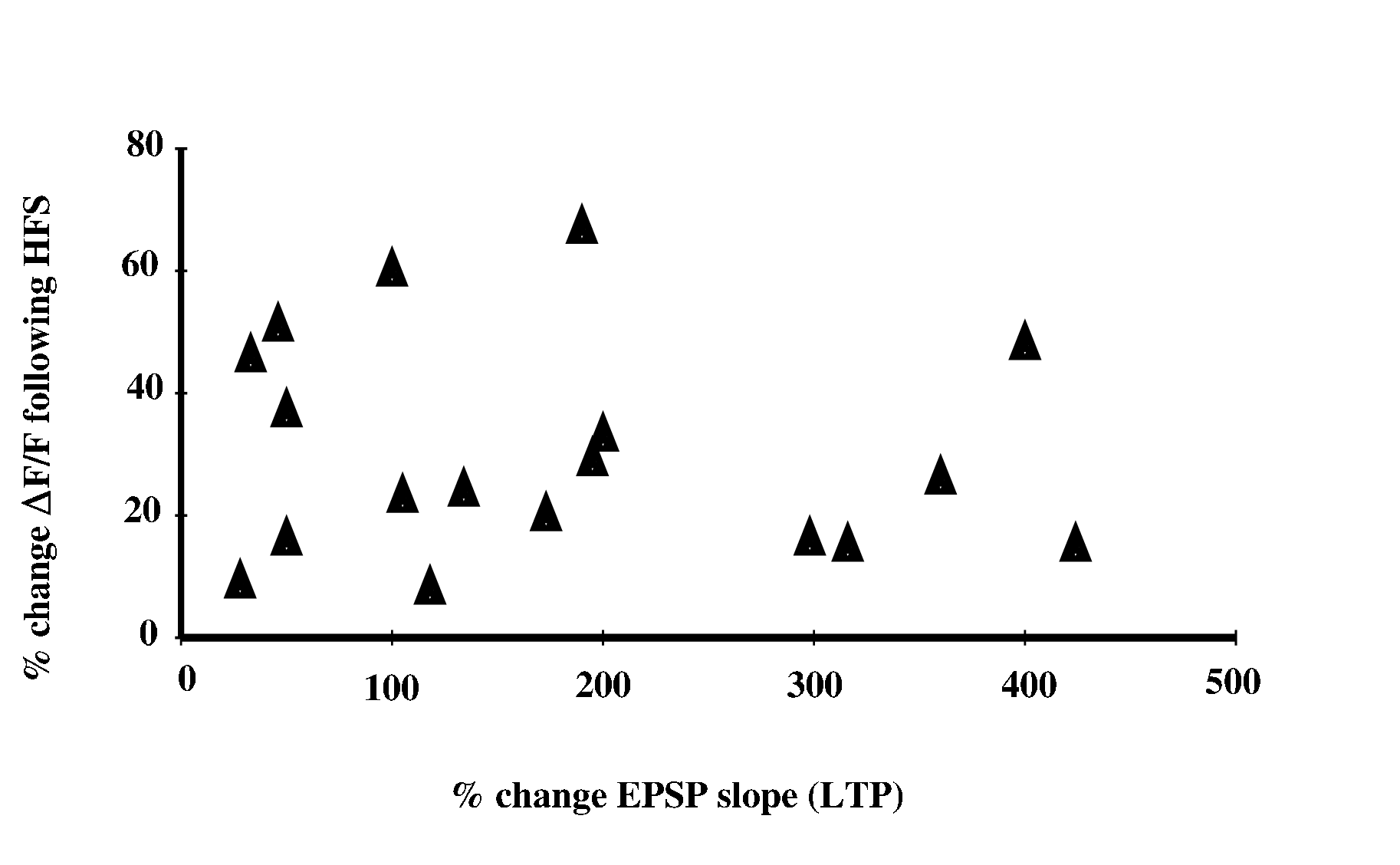}}

\caption[  Increases in dendritic Ca$^{2+}$ signals at the synapse following 100 Hz pairing stimulation as a function of percent LTP for each experiment (30 min post-HFS).]
{{\bf   Increases in dendritic Ca$^{2+}$ signals at the synapse following 100 Hz pairing stimulation as a function of percent LTP for each experiment (30 min post-HFS).} {There appears to be no correlation between the size of the increase in dendritic Ca$^{2+}$ signal and the size of the LTP obtained.}}
\label{Fig24}
\end{figure}

\chapter*{4 Discussion} \setcounter{chapter}{4}
\addcontentsline{toc}{chapter}{4 Discussion} \setcounter{section}{0}
\setcounter{subsection}{0} \setcounter{subsubsection}{0}

\section {Temperature and external Ca$^{2+}$-dependence of LTD}

The present study indicates that bath temperature does not dramatically influence the degree of depression observed following low-frequency stimulation.  Equivalent LTD was produced at both 23 and 32$^\circ$C.  In contrast, the extracellular bath concentration of Ca$^{2+}$ had marked effects on LTD induction.  When a Ca$^{2+}$-free solution was washed in before the application of the induction protocol, synaptic transmission was reduced significantly, and no EPSPs were recorded during the low-frequency stimulation.  Although it is unclear whether the lack of LTD was due to a failure to evoke transmitter release or to induce a postsynaptic Ca$^{2+}$ influx, low-frequency stimulus-induced homosynaptic LTD does not appear to be the result of some stimulation-induced pathology, because responses fully recovered after a return to normal medium.  The lack of LTD in  high-Ca$^{2+}$ medium is also difficult to interpret, because this may reflect changes in the amount of transmitter being released, an increase in postsynaptic Ca$^{2+}$, or an alteration in membrane protein function due to changes in charge screening.  Again, prolonged low-frequency stimulation did not generate LTD in this situation, indicating that LTD is not dependent on stimulation alone.  Furthermore, responses were depressed when the low-frequency stimulation was administered in normal medium.

\section {VGCC-dependence of LTD}

Our data indicate that Ca$^{2+}$ influx via at least two subtypes of VGCC appears to play an important role in the induction of homosynaptic LTD in the CA1 region in vitro.  The fact that the L-type VGCC antagonist nimodipine prevented the induction of homosynaptic LTD also supports the findings of Bolshakov and Siegelbaum (1994), who prevented homosynaptic LTD induction in slices obtained from very young animals with the L-type VGCC antagonist nitrendipine.  It remains unclear why a third L-type antagonist, nifedipine, has been unsuccessful in preventing LTD induction in animals of similar age [. mulkey malenka 1992 @11104@, selig bear 1995 @13073@.].  This may reflect a difference in the experimental methodology or a difference in the site of action of these antagonists.  In addition to the contribution of L-type channels to homosynaptic LTD, it also appears that either or both the R and T subtypes of VGCC contribute to the induction of homosynaptic LTD.  These channels are richly distributed in the apical dendrites of CA1 pyramidal neurons [.christie eliot 1995 @12874@, magee science 1995 @12705@.], where they are able to provide a source of Ca$^{2+}$ in response to both weak and strong synaptic stimulation [.magee christie 1995 @13563@.]. Recent evidence from our laboratory indicates that stimuli that are subthreshold for the generation of action potentials are insufficient to generate LTD in individual neurons [.christie eliot 1995 @12874@.].  This would indicate that it is more likely that R- rather than T-type VGCCs are active in homosynaptic LTD.  Fluorescence imaging studies indicate that Ca$^{2+}$ influx via high-threshold L- and R-type VGCCs occurs only when somatic action potentials are generated [.christie eliot 1995 @12874@, magee christie 1995 @13563@.], suggesting a role for these channels in LTD. 

It seems possible that the involvement of the NMDA receptor might be to prolong the period of postsynaptic depolarization so that VGCCs are more likely to be activated and LTD induced.  Hell and colleagues (1996) have in fact shown that NMDA receptor activation in CA1 cells may persistently increase Ca$^{2+}$ influx through at least L-type channels after intense synaptic activity (see also Chetkovich et al., 1991).  It is interesting that homosynaptic LTD, like LTP in this region, is susceptible to manipulations that alter the levels of postsynaptic Ca$^{2+}$ in the dendrites and somata of CA1 pyramidal cells.  Both LTP and LTD have been prevented in the CA1 region by antagonists of NMDA receptors [.collingridge schaffer 1983 @5376@, dudek bear 1992 @10909@.], L-type VGCC antagonists [.bolshakov science 1994 @11961@, grover nature 1990 @9815@.], and now by the R- and T-type VGCC antagonist NiCl$_{2}$ [.ito miura 1995 @13613@.].  The mechanism by which these channels exert their effects on LTD induction remains unclear, however.  It is possible that LTD induction requires a particular level of postsynaptic Ca$^{2+}$ [.lisman 1994 tins @12202@.] and that any manipulations that substantially reduce postsynaptic Ca$^{2+}$ influx prevent the induction of LTD, although it is likely that the manner by which this Ca$^{2+}$ enters the cell is also a critical factor in LTD induction.

\section{Postsynaptic Ca$^{2+}$ imaging during LTD/LTP induction}

 These studies provide support for the hypothesis that synaptic plasticity depends on postsynaptic  [Ca$^{2+}$] and frequency of stimulation, such that low frequencies produced modest increases in postsynaptic [Ca$^{2+}$] and LTD, while higher frequencies produced larger increases in postsynaptic [Ca$^{2+}$] and LTP.  Application of nimodipine or APV reduced the stimulation-induced increases in postsynaptic Ca$^{2+}$, necessitating the application of higher frequency stimulation to achieve a given degree of plasticity.  These reductions in postsynaptic [Ca$^{2+}$] appear to make the induction of LTD more favorable than the induction of LTP, shifting the LTD/LTP transition point toward higher stimulus frequencies.

\subsection{Frequency dependence of LTD/LTP induction}

	In these experiments low-frequency stimulation (3 and 10~Hz) reliably resulted in LTD while higher frequency stimulation (30~Hz and above) resulted in LTP.  Through the use of fluorescence imaging, we were able to determine, for several locations in the neuron, the size of the change in postsynaptic [Ca$^{2+}$] for each of the above stimulus frequencies and to correlate these Ca$^{2+}$ signals with the amount of plasticity obtained.  As predicted by the Lisman model (1989), we observed that postsynaptic [Ca$^{2+}$] increased with increasing stimulus frequency, as well as with the increasing likelihood of LTP induction.  The data obtained in the control condition allowed us to determine a range of postsynaptic [Ca$^{2+}$] above which we would expect to see LTP and below which we would expect to see LTD.  We then used this information to attempt to shift the threshold for LTD/LTP induction toward higher stimulus frequencies by reducing NMDA receptor- and VGCC-mediated Ca$^{2+}$ influx. 

\subsection{Role of L-type VGCCs}

It has been shown that dendritic voltage-gated Ca$^{2+}$ entry within 150~$\mu$m of the soma of CA1 pyramidal neurons is primarily due to the activity of L-type VGCCs [.Westenbroek Nature 1990 @10128@, Magee Physiol 1995, Christie Ito 1995 @12874@, Jaffe nature 1992 @10990@.].  We used 10~$\mu$M nimodipine to block these channels, resulting in reductions in the rise in postsynaptic [Ca$^{2+}$] of about 30\% in the soma and the proximal and middle dendritic regions of the neuron at all frequencies tested.  This corresponded to a blockade of LTD induction with 3 Hz stimulation but not 10~Hz stimulation, and resulted in induction of LTD rather than LTP with 30, 50 and 100~Hz stimulation.  No plasticity was observed with 200~Hz stimulation, which seems to suggest a shift in the threshold for LTP induction to higher levels of [Ca$^{2+}$]$_{i}$  than could be obtained with 200~Hz.     

\subsection{Role of NMDA receptor mediated Ca$^{2+}$ influx}

	Application of 10~$\mu$M APV, a low concentration intended to reduce but not completely block NMDA receptor-mediated Ca$^{2+}$ influx, also appeared to reduce the rise in postsynaptic [Ca$^{2+}$] by about 30\% in the proximal and middle dendritic regions at all frequencies tested, but did not appear to reduce [Ca$^{2+}$]$_{i}$ in the soma, probably due to the absence of synaptically activated NMDA receptors located in this region.  It is possible that the blockade of NMDA receptors reduces the amount of depolarization during the tetanus, reducing the activation of the high-voltage activated L-type channels and decreasing the Ca$^{2+}$ influx via these VGCCs [.Miyakawa Neuron 1992 @11090@.].   Although tetanus-induced depolarizations in the soma during APV application were not reduced relative to controls, we cannot rule out reductions in the dendrites.  In the presence of 10~$\mu$M APV, no LTD was induced with 3~Hz or 10~Hz stimulation, but LTD rather than LTP was induced with 30, 50, and 100~Hz stimulation.  A reduced amount of LTP was induced with 200~Hz stimulation in the presence of 10~$\mu$M APV.  This LTP seemed to exhibit a delayed onset similar to the NMDA independent LTP induced by 200~Hz stimulation that was observed by Grover and Teyler (1990).    
 
 	It should be noted that in the presence of 50~$\mu$M APV, a concentration intended to effectively block NMDA receptor-mediated Ca$^{2+}$ influx, no long-term plasticity was observed at any frequency, but at 200~Hz a slight, transient depression was observed that recovered before the end of the 30 min recording period.  This result differs from the potentiation observed by Grover and Teyler (1990) with 200 Hz stimulation.  It is possible that this difference is due to their use of older animals (>6 weeks) in which non-NMDA LTP may be more easily induced than in the younger animals used here [.Shankar 1998 @15670@.].  Our results in 50~$\mu$M APV seem to suggest an extreme shift to the right of the LTP/LTD induction curve, in which Ca$^{2+}$ influx is reduced to such an extent that no frequency of stimulation is sufficient to evoke the level of [Ca$^{2+}$]$_{i}$ necessary to induce LTD.  In the presence of 200 Hz stimulation, however, previously shown to be the frequency at which voltage-gated Ca$^{2+}$ influx plays a stronger role in the induction of plasticity [.Grover Teyler nature 1990 @9815@.], it is possible that some VGCC-mediated Ca$^{2+}$ influx occurs to account for the small amount of transient depression observed.  Although 50~$\mu$M APV caused a much stronger reduction in plasticity at all frequencies than 10~$\mu$M APV, the reductions in postsynaptic [Ca$^{2+}$] in the presence of 50~$\mu$M APV were only slightly greater than those observed in the presence of 10~$\mu$M APV.  This may be due to the fact that the additional Ca$^{2+}$ influx blocked by 50~$\mu$M APV belonged to a pool of Ca$^{2+}$ that could not be measured with our methods, for example, at the dendritic spines [.Schiller 1998 @16012@, Yuste Denk 1995 @12893@.].  

\subsection{Regional differences in Ca$^{2+}$ dependence of LTD/LTP}

	When the above changes in synaptic strength are plotted as a function of postsynaptic [Ca$^{2+}$] for all experiments, there is an obvious difference in the relationship between plasticity and [Ca$^{2+}$]$_{i}$ in the soma versus the dendrites.  For each region the trend is similar, with the largest postsynaptic Ca$^{2+}$ signals corresponding to potentiation, intermediate values corresponding to depression, and the smallest signals corresponding to no observed plasticity.  The signals in the soma that corresponded to potentiation, however, were larger than those in the dendrites that corresponded to potentiation.  Hence, the apparent [Ca$^{2+}$]$_{i}$ threshold for LTP induction was higher in the soma than in the dendrites.  This may be due to the fact that, for any given experiment, the Ca$^{2+}$ signals were largest in the soma, probably owing to the direct stimulation of the soma in the pairing protocol, and does not necessarily indicate different Ca$^{2+}$ thresholds for the induction of plasticity in different regions of the neuron.  Also, the relationship between plasticity and [Ca$^{2+}$]$_{i}$ in the presence of APV is not the same for the soma as in the dendrites. In the presence of APV, plasticity was reduced or blocked though there were no reductions in somatic [Ca$^{2+}$]$_{i}$,  When these APV data are plotted separately (top panel in Fig. 7), the remainder of the data for the soma show a similar relationship as in the dendrites, where low [Ca$^{2+}$]$_{i}$ leads to depression and high [Ca$^{2+}$]$_{i}$ leads to potentiation.  Since APV blocks plasticity without causing a reduction in somatic Ca$^{2+}$ influx, it seems reasonable to conclude that increases in dendritic [Ca$^{2+}$]$_{i}$ are more relevant for the induction of plasticity than those in the soma.  Also, as excitatory synapses are located in the dendrites, this would be the most likely region for postsynaptic [Ca$^{2+}$] to have its effects on the induction of plasticity.  Our data do not allow us to adequately address  the relative roles of the Ca$^{2+}$ influx in each region, though there may be important differences in their contributions to plasticity.

\subsection{Conclusions and interpretation}

	While others have demonstrated a shift of the threshold for LTD/LTP induction achieved by modulations of postsynaptic [Ca$^{2+}$] [.Artola Nature 1990 @9647@, Cummings Neuron 1996 @13586@, Kimura 1990, Brocher 1992, Mulkey Malenka 1992 @11104@.], we were able to use fluorescence imaging to simultaneously measure the changes in postsynaptic [Ca$^{2+}$] that corresponded to these shifts in threshold and to correlate postsynaptic [Ca$^{2+}$] with plasticity over the entire range of commonly used stimulus frequencies.  Hansel and colleagues (1997)  have achieved similar results for three common stimulus frequencies in imaging studies in neurons of the visual cortex.  It should be noted that we chose to measure postsynaptic [Ca$^{2+}$] at the end of the first epoch of 100 pulses rather than at the end of an uninterrupted train of 900 or 400 pulses, which might seem to better reflect the cumulative increase in [Ca$^{2+}$]$_{i}$ during the entire protocol.  For all frequencies of stimulation, however, postsynaptic [Ca$^{2+}$] levels usually rose to a maximum by the end of the first 25 of the 100 pulses, and did not vary significantly between multiple epochs.  Our signals returned to baseline between epochs and showed no increase with repeated stimulation, unlike those observed by Hansel and colleagues (1997).  Hansel and colleagues also reported significantly higher peak amplitudes at dendritic locations, while our largest amplitude signals occurred in the region of the soma.  This may be due to our direct depolarization of the soma in the pairing stimulus protocol, which probably evokes large increases in somatic [Ca$^{2+}$] not evoked by synaptic stimulation alone.  Although not addressed in this study, the rise and decay times of these Ca$^{2+}$ signals may also be important for the induction of plasticity.  In addition to VGCCs and NMDA receptors, there are likely to be contributions from other sources of postsynaptic Ca$^{2+}$, such as release from intracellular stores following metabotropic glutamate receptor activation [.frenguelli dendrites 1993 @11463@, Jaffe waves 1994 @12139@, Linden Smeyne 1994 @12479@, kawabata 1996 @13770@.].

	It should also be noted that Neveu and Zucker failed to show such a relationship between postsynaptic [Ca$^{2+}$] and the polarity of synaptic plasticity in experiments where postsynaptic [Ca$^{2+}$] was manipulated by photolysis of caged Ca$^{2+}$ using the compound nitr-5 [.neveu neuron 1996 @15212@.].  It is likely that, due to the limitations of the caged Ca$^{2+}$ compound used, the temporal profile of the Ca$^{2+}$  increases evoked by caged Ca$^{2+}$ did not replicate those normally achieved during plasticity-inducing stimulation or did not reach sufficient levels in the postsynaptic cell.  In a more recent study conducted by this group, however, the use of a newer caged Ca$^{2+}$ compound, nitrophenyl EGTA, allowed both larger increases and more prolonged rises in postsynaptic Ca$^{2+}$, and may have more closely replicated the changes in postsynaptic Ca$^{2+}$ that occur during the induction of LTP and LTD, respectively [.Yang Tang 1999 @16838@.].  In this new study, it was found that prolonged modest rises in Ca$^{2+}$ reliably induced LTD, while brief increases of a high magnitude induced LTP.  Although postsynaptic Ca$^{2+}$ was not measured in their study, their data indicate that the magnitude of increases in postsynaptic Ca$^{2+}$ are predictive of the direction of change in synaptic strength.               

	Our results suggest that, at least in the 2-3 week animals we used,  both voltage-gated Ca$^{2+}$ channels and NMDA receptors contribute to the Ca$^{2+}$ influx necessary for the induction of synaptic plasticity.  While there are some differences between the effects seen with APV and nimodipine, the data seem to indicate that, in general, the blockade of postsynaptic Ca$^{2+}$ influx through either of these channel types leads to a reversal of LTD as well as of LTP.  Thus, we hypothesize that, if high levels of postsynaptic [Ca$^{2+}$] lead to kinase activation and LTP and low levels lead to phosphatase activation and LTD, then with high-frequency stimulation these blockers reduce the [Ca$^{2+}$]$_{i}$ to a level that is no longer sufficient to activate the relevant kinases but which is still sufficient to activate the phosphatases that lead to LTD.  Conversely, in the cases of low-frequency stimulation,  the blockers reduce the [Ca$^{2+}$]$_{i}$ to such a low level that it is no longer sufficient to activate even the phosphatases, leading to a blockade of LTD.  In addition, our data, not surprisingly, suggest that increases in dendritic [Ca$^{2+}$] are more predictive of the magnitude and direction of plasticity than those in the soma.  These data provide support for the hypothesis that there is a [Ca$^{2+}$]$_{i}$ threshold for the induction of LTD and LTP such that the magnitude and direction of changes in synaptic strength depend on the level of postsynaptic [Ca$^{2+}$] achieved during stimulation.  Thus, increases in postsynaptic [Ca$^{2+}$]  play an important role in translating different frequencies of stimulation into changes in synaptic strength.

\section{Ca$^{2+}$ Imaging Reveals Localized Increases in Dendritic Action Potential Amplitude Following LTP Induction}

	It has been shown that the apical dendrites of CA1 hippocampal neurons contain a large density of transient or A-type K$^{+}$ channels, which appear to play a role in modulating EPSPs and back-propagating action potentials, resulting in regulation of dendritic signal propagation [.Hoffman nature @15041@.]. It has also been shown that selective activators of protein kinases shift the activation curves of these dendritic K$^{+}$ channels and decrease their activity, resulting in increased action potential amplitude [.Hoffman PKA @15975@.]. As it is well-known that LTP induction causes such an activation of kinases, it seemed likely that increases in the amplitude of back-propagating action potentials might occur following LTP induction.  Preliminary experiments employing whole-cell recordings from the dendrites before and after the induction of LTP did not reveal any such increases in back-propagating action potential amplitude (Hoffman and Johnston, unpublished results).  However, if the increases in action potential amplitude were localized to the area of the synaptic input, it is unlikely that the dendritic electrode was located at the precise site of the synapse where the effect would most likely occur and where a difference could be measured.  In addition, it is possible that dialysis of intracellular mediators at the dendritic recording site might prevent localized biochemical reactions necessary for phosphorylation of K$^{+}$ channels and increased action potential amplitude at the synapse.     

	We thus hypothesized that LTP induction does indeed result in localized decreases in dendritic K$^{+}$ channel activity and increases in amplitude of back-propagating action potentials, and proposed to observe these increases in amplitude indirectly as localized increases in dendritic Ca$^{2+}$signals following LTP induction. With the use of the fluorescent Ca$^{2+}$ indicator fura-2, it was not necessary to record at the precise site of the synapse to detect the increases in action potential amplitude and, because the whole-cell electrode was located at the soma, far from the site of the synaptic input, dialysis of intracellular mediators at the synapse was less of a concern.  

\subsection{ Increases in dendritic Ca$^{2+}$ signals following LTP}

	A brief train of action potentials was applied before the application of HFS, 5 min after the application of HFS, and 30 min after the application of HFS solely for the purposes of measuring postsynaptic Ca$^{2+}$ signals.  At the 5 min time point, consistent localized increases in the size of dendritic Ca$^{2+}$signals were observed after the induction of LTP in 8 out of 8 experiments.  The observed increases from the 5 min time point did not uniformly persist until the 30 min time point, nor did any further increases appear to occur.  This may indicate that the changes in Ca$^{2+}$ signals we observe are related more to the induction of LTP than to the expression of LTP.  It is also possible that, since the dendritic Ca$^{2+}$ signals became more variable at the 30 min time point, our inability to detect any uniform changes at 30 min may be due to a limitation of our ability to reliably record Ca$^{2+}$ signals at this late point in the experiment, perhaps a result of decreased cell health or increased background fluorescence.

\subsection{Synaptic localization of increases in dendritic Ca$^{2+}$}

It was possible to determine at the beginning of each experiment the location of the synaptic input by observing the location of the small Ca$^{2+}$ signal evoked in response to a short train of subthreshold EPSPs [. Magee subthreshold @13563@.].  Subsequently when the increases in dendritic Ca$^{2+}$ signals were observed following LTP, it was possible to determine whether they occurred at the site of the synaptic input or at some other location in the dendrites.  In all 8 of the experiments where LTP was induced, the post-HFS increases in dendritic Ca$^{2+}$ signals were observed in the 25 $\mu$m segment of the dendrite shown to correspond to the site of synaptic input, or in the 25 $\mu$m segment adjacent to it.  Thus it appears that the increases in dendritic Ca$^{2+}$ signals following LTP occur by a mechanism that is localized to the synapse.

\subsection {Dependence on synaptic input}

It has been shown that back-propagating action potentials alone do not produce LTP in the absence of synaptic stimulation [.magee Hebbian @14316@.].  It was of interest to determine whether the observed localized increases in the size of dendritic Ca$^{2+}$ signals following 100 Hz pairing stimulation were due to the presence of back-propagating action potentials alone or whether they required the presence of synaptic input.  Thus, the above experiments were repeated with 100 Hz back-propagating action potentials only and, as expected, no LTP was induced.  In addition, there were no observed increases in the size of the dendritic Ca$^{2+}$ signals in any region of the dendrite following the stimulation.  It appears, then, that overall strong depolarization of the dendritic tree is not sufficient to activate whatever mechanisms are responsible for the localized increases in dendritic Ca$^{2+}$ signals and presumably action potential amplitude.  Rather, it appears that there is a need for synaptic input, and possibly for NMDA receptor activation.  

\subsection {Dependence on NMDA receptor activation}

To determine whether NMDA receptor activation was necessary for the localized increases in the size of dendritic Ca$^{2+}$ signals following HFS, experiments were performed in which 100 Hz pairing stimulation was applied in the presence of 50~$\mu$M APV, a concentration known to reliably block LTP induction.  As expected, no LTP was observed and, in addition, no increases in the size of the dendritic Ca$^{2+}$ signals were observed in any region of the dendrite following the stimulation.  This may indicate a need for NMDA receptor activity for the activation of the mechanisms necessary to cause localized increases in dendritic Ca$^{2+}$ signals.  Alternatively, it may be that the lack of observed increases in dendritic Ca$^{2+}$ signals is due to the fact that no LTP was induced, and that some biochemical step resulting from LTP induction was lacking when APV was used to block LTP.  

\subsection {Dependence on protein kinase activity}

It is known that protein kinase activation is necessary for the induction of LTP, and that LTP can be blocked by the addition of inhibitors of various kinases.  In addition, activators of cAMP-dependent protein kinase (PKA) and protein kinase C (PKC) have been found to increase the amplitude of EPSPs and population spikes in hippocampal neurons [.slack efficacy @14547@,hu albert @15164@.], as well as to decrease the activity of dendritic K$^{+}$ channels and to increase the amplitude of back-propagating action potentials [.hoffman downregulation @15975@.].  Thus, our hypothesis for the cause of the observed dendritic Ca$^{2+}$ increases involves the phosphorylation of dendritic K$^{+}$ channels by various kinases and the shifting of their activation curves to decrease their activity.  It was of interest, then, to attempt to block the induction of LTP with inhibitors of protein kinases to determine whether localized increases in dendritic Ca$^{2+}$ signals would still occur following HFS.  This would also determine whether the APV block of the increase in Ca$^{2+}$ signals noted above was due to the absence of NMDA receptor activity or, rather, to the lack of kinase activity resulting from blockade of LTP induction.  We used the broad spectrum protein kinase inhibitor H7 at a concentration of 300~$\mu$M which has been shown to reliably block LTP induction [.Klann enhanced @14823@, malinow tsien kinase @8805@.].  As expected, no LTP was observed in the presence of H7 and, in addition, there were no observed increases in the size of dendritic Ca$^{2+}$ signals following HFS.  This seems to indicate a need for the activity of one or more kinases to give rise to the observed increases in dendritic Ca$^{2+}$ signals.     

	As H7 acts by inhibiting the activity of several kinases, it was of interest to further elucidate the role of a particular protein kinase in this phenomenon.  We used the compound H89, an inhibitor of PKA, at a concentration of 10~$\mu$M [.roberson binding @16970@.]. Studies have shown PKA activity to be necessary for the late phase of LTP (L-LTP) which requires three or more trains of tetanic stimuli separated by at least 5 min and which occurs 1 to 3 hours after stimulation.  However, inhibitors of PKA have been shown to have only small effects on early LTP (E-LTP) which is the type induced by a single train of tetanic stimuli and which lasts 1 to 3 hours [.frey huang @11464@.].  As it is the early phase of LTP that we observe in these experiments, it is not surprising that H89 did not block the induction of LTP.  Surprisingly however, when LTP was induced in the presence of H89, we did not observe the localized increases in dendritic Ca$^{2+}$ signals that were observed in experiments where LTP was induced under normal conditions.       

	It appears, then, that PKA is necessary to give rise to the localized increases in dendritic Ca$^{2+}$, although it is not necessary for the induction of the early phase of LTP.  In all of the previous experiments, increases in dendritic Ca$^{2+}$ signals were observed whenever LTP was induced, and absent in all experiments in which LTP was blocked.  Interestingly, it seems that PKA may have a somewhat different role in this phenomenon.  It is likely that in our experiments the actions of PKA take place very early in the induction of E-LTP, as the effects we measure are present only 5 min after application of HFS.  It is possible that PKA acts as more of an amplification signal at this early stage in LTP induction and that, while it may phosphorylate K$^{+}$ channels to result in their downregulation and a localized increase in neuronal excitability and Ca$^{2+}$ influx, this may serve to enhance the induction of LTP but may not be absolutely necessary for its induction.  For example, as it is known that induction of L-LTP requires multiple trains of tetanic stimuli separated by at least 5 min, such actions of PKA may result in short-term increases in Ca$^{2+}$ signals in the synaptic region that give rise to amplification during the subsequent tetani used to induce L-LTP.  Interestingly, activators of PKA have also been shown to result in the increase of VGCC activity [.chetkovich gray @13114@.], another action which would serve to amplify Ca$^{2+}$ signals leading to LTP.    

\subsection {Conclusions and interpretation}

We conclude from these data that the HFS used to induce LTP results in localized increases in dendritic Ca$^{2+}$ influx which may represent a localized increase in action potential amplitude at or near the synapse.  We propose that this occurs through the activation of one or more kinases which act locally at the synapse to phosphorylate dendritic K$^{+}$ channels and to decrease their activity.  We cannot rule out, however, a synapse-specific increase in the activity of VGCCs due to their phosphorylation.  In order to distinguish between these two possibilities, it would be useful to be able to record action potentials at the synapse to determine whether they are indeed increased following LTP, or to record from K$^{+}$ channels located at the synapse to detect changes in their activity following LTP.  In addition, the development of antibodies to the various phosphorylation sites on the Kv4.2 K$^{+}$ channel, currently in progress in David Sweatt's laboratory, would make it possible to determine whether there was indeed phosphorylation of these channels of interest and whether it occurred in a synapse-specific fashion.

\chapter*{5 Appendix} \setcounter{chapter}{5}
\addcontentsline{toc}{chapter}{5 Appendix} \setcounter{section}{0}
\setcounter{subsection}{0} \setcounter{subsubsection}{0}

\section {Increased CaMKII autophosphorylation following LTP induction in individual CA1 pyramidal neurons}

Ca$^{2+}$/calmodulin dependent protein kinase II (CaMKII) is thought to play a role in synaptic plasticity in the CA1 region of the hippocampus.  The high-frequency stimulation used to induce LTP is known to result in substantial increases in intracellular Ca$^{2+}$ in the postsynaptic cell.  CaMKII is activated in the presence of Ca$^{2+}$ and calmodulin and becomes autophosphorylated at threonine-286.  It then becomes autonomously active, even in the absence of Ca$^{2+}$, until it is dephosphorylated by cellular phosphatases [.miller kennedy cell @7631@.].         

	Mary Kennedy and colleagues have developed a semiquantitative immunohistochemical method for  visualizing autophosphorylated CaMKII (P-CaMKII) in fixed hippocampal slices.  Their method utilizes a phosphosite-specific monoclonal antibody that recognizes CaMKII only when it is phosphorylated at threonine-286.  Extracellular recordings were used to show that substantial increases in autophosphorylation of CaMKII occur in dendrites and somas in CA1 hippocampus within 1mm of the stimulating electrode 30 min after tetanization of the Schaffer collaterals [.ouyang @16148@.]. 

	These data indicate that strong tetanization of the Schaffer collaterals causes substantial increases in autophosphorylation of CaMKII in dendrites and somas in CA1 hippocampus within 1mm of the stimulating electrode. 
Thus, we hypothesized that in whole-cell experiments P-CaMKII would be increased selectively in the individual cell that received LTP-inducing stimulation consisting of subthreshold synaptic stimulation paired with back-propagating action potentials.   

	To test this hypothesis,  whole-cell recordings were performed from CA1 neurons using an internal solution that contained the dye biocytin to mark the cell of interest.  100 Hz pairing stimulation was applied to induce LTP in the cell of interest only, such that surrounding cells received only subthreshold stimulation.  Subthreshold synaptic stimulation was delivered with an extracellular stimulating electrode placed in the stratum radiatum.  These stimuli were paired with back-propagating action potentials induced by current pulses applied at the soma with a whole-cell electrode.  It has been shown previously that LTP is induced only in cells that receive both subthreshold synaptic stimulation and back-propagating action potentials, but not in cells that receive only subthreshold synaptic stimulation [.magee hebbian @14316@.].  
At the end of each experiment, the slices were fixed and shipped on ice overnight to the Kennedy lab at Caltech.  There the Kennedy lab used their phosphosite specific antibody to detect increases in P-CaMKII in the biocytin filled cell of interest.  

	The histology was performed by Susan Catalano of the Kennedy Lab.  When the fixed slices were received, 50 $\mu$m sections were cut and the slices were incubated with primary antibody (mouse anti-P-CaMKII) then with secondary antibody (anti-mouse conjugated to the fluorescent dye Cy-3).  In order to identify the biocytin filled cell of interest, biocytin was visualized with Neuravidin plus a different fluorescent dye, Oregon Green.  These two fluorescent dyes were able to be visualized individually at different wavelengths. 
  
	Out of all slices sent for histology, cells chosen for analysis, included 9 cells that were stimulated and underwent LTP, and 4 cells that were stimulated in the presence of  50 uM APV and that showed no LTP. Of the 9 cells that underwent LTP, 5 showed an increase in P-CaMKII.  Of the 4 cells stimulated in the presence of APV and that showed no LTP, all showed an increase in P-CaMKII.  An additional 8 cells were recorded from and filled with biocytin, but received no 100 Hz stimulation, and none showed increased P-CaMKII.   

	It appears, then, that when LTP-inducing stimulation is limited to a single CA1 neuron, that neuron alone often shows increased P-CaMKII, suggesting that LTP induction results in increased CaMKII activity in only the stimulated neuron.  These increases do not appear to be an artifact of whole cell recording, as none of the unstimulated, biocytin-filled controls showed increases in P-CaMKII.  In addition, our data seem to indicate that the 100 Hz pairing stimulation, even in the presence of APV is sufficient to evoke the observed increases in P-CaMKII.  

	It should be noted that extracellular recordings performed previously by the Kennedy lab showed that APV totally blocked increases in P-CaMKII [.ouyang @16148@.].  Our data do not agree with this finding.  It is likely that our protocol of delivering back-propagating action potentials via the somatic whole-cell electrode allows substantial increases in Ca$^{2+}$ even in the presence of APV, perhaps leading to increases in P-CaMKII, whereas in their extracellular experiments APV may have been sufficient to block most action potentials and Ca$^{2+}$ increases evoked by field stimulation.         
Thus, although our findings seem to indicate that increases in P-CaMKII are associated with the stimulation used to induce LTP, they contribute no evidence that autophosphorylation of CaMKII is a process necessary or sufficient for the induction of LTP.

\chapter*{6 Summary and Significance}
\setcounter{chapter}{6}
\addcontentsline{toc}{chapter}{6 Summary and Significance}
\setcounter{section}{0}

\newcounter{l10}

\begin{list}{\arabic{l10}.}{\usecounter{l10}}

\item The magnitude of LTD induction does not vary with changes in bath temperature from 23$^\circ$C to 32--35$^\circ$C, indicating that homosynaptic LTD induction is not as temperature-sensitive as LTP.  

\item  LTD induction with 3 Hz stimulation is sensitive to extracellular Ca$^{2+}$ concentration.  LTD is prevented with decreased extracellular Ca$^{2+}$ (0 mM), and in 4mM Ca$^{2+}$ 3 Hz stimulation results in slight potentiation.  

\item  LTD induction with 1 or 3 Hz stimulation is dependent on both T- and L-type voltage-gated Ca$^{2+}$ channel activity, as it is blocked by either 10~$\mu$M nimodipine or 25~$\mu$M NiCl$_{2}$. 

\item Fluorescence imaging experiments demonstrated that the level of postsynaptic [Ca$^{2+}$] observed during the stimulation used to induce LTD or LTP varies with frequency of stimulation and with the magnitude and direction of change in synaptic strength.  

\item LTD was achieved with stimulation from 3 to 10 Hz while LTP was achieved with stimulation of 30 Hz and above.

\item Although there was no precise magnitude of postsynaptic [Ca$^{2+}$] signal that corresponded to the division between LTP and LTD, there was a range of Ca$^{2+}$ values above which LTP was observed and below which LTD was observed.  

\item  10~$\mu$M nimodipine blocked Ca$^{2+}$ signals by about 30\% in the soma and dendrites at each stimulus frequency and shifted the crossover point of the plasticity vs. frequency curve toward higher frequencies of stimulation, such that frequencies that normally resulted in LTP resulted in LTD, and frequencies that normally resulted in LTD gave rise to no plasticity. 

\item 10~$\mu$M APV blocked Ca$^{2+}$ signals by about 30\% in the proximal and middle dendrite at each stimulus frequency but had no effect on somatic Ca$^{2+}$ signals. Like nimodipine, it shifted the crossover point of the plasticity vs. frequency curve toward higher frequencies of stimulation.  

\item  50~$\mu$M APV, like 10~$\mu$M APV, blocked Ca$^{2+}$ signals by about 30\% in the proximal and middle dendrite at each stimulus frequency but had no effect on somatic Ca$^{2+}$ signals.  No plasticity was obtained in the presence of this higher concentration of APV at any frequency tested.

\item  In these experiments both voltage-gated Ca$^{2+}$ channels and NMDA receptors appear to contribute to the Ca$^{2+}$ influx necessary for the induction of synaptic plasticity. 

\item  Since APV blocks plasticity without causing a reduction in somatic Ca$^{2+}$ influx, it appears that increases in dendritic [Ca$^{2+}$]$_{i}$ are more relevant for the induction of plasticity than those in the soma.    

\item   There appears to be a [Ca$^{2+}$]$_{i}$ threshold for the induction of LTD and LTP such that the magnitude and direction of changes in synaptic strength depend on the level of postsynaptic [Ca$^{2+}$] achieved during stimulation.

\item   Application of 100 Hz pairing stimulation resulted in LTP induction and in  consistent localized increases in the size of dendritic Ca$^{2+}$signals 5 min after stimulation. 

\item  It was possible to determine the location of the synaptic input at the beginning of each experiment by observing the location of the small Ca$^{2+}$ signal evoked in response to a short train of subthreshold EPSPs.

\item In experiments where LTP was induced, the increases in dendritic Ca$^{2+}$ signals were observed in the 25 $\mu$m segment of the dendrite shown to correspond to the site of synaptic input, or in the 25 $\mu$m segment adjacent to it.

\item Overall strong depolarization of the dendritic tree is not sufficient to activate the mechanisms responsible for the localized increases in dendritic Ca$^{2+}$ signals.  Rather, it appears that there is also a need for synaptic input.    

\item   50 $\mu$M APV blocked LTP induction as well as the increases in the size of the dendritic Ca$^{2+}$ signals. 

\item  No LTP was obtained and no increase in the size of the dendritic Ca$^{2+}$ signals was observed in the presence of 300~$\mu$M H7.  This may indicate a need for the activity of one or more kinases to give rise to the observed increases in dendritic Ca$^{2+}$ signals.

\item  10~$\mu$M H89 did not block the induction of LTP but blocked the localized increases in dendritic Ca$^{2+}$ signals normally observed following LTP.  It appears that PKA is necessary to give rise to the localized increases in dendritic Ca$^{2+}$, although it is not necessary for the induction of the early phase of LTP.

\item  The high-frequency stimulation used to induce LTP results in localized increases in dendritic Ca$^{2+}$ influx from back-propagating action potentials which may represent a localized increase in action potential amplitude at or near the synapse.  We propose that this occurs through the activation of one or more kinases which act locally at the synapse to phosphorylate dendritic K$^{+}$ channels and to decrease their activity.

\end{list}

\newpage

\addcontentsline{toc}{chapter}{References}

\end{document}